\documentclass[prd,twocolumn,a4paper,superscriptaddress,floatfix]{revtex4}
\usepackage{graphicx}
\begin{document}

\newcommand{\be}{\begin{equation}}
\newcommand{\ee}{\end{equation}}
\newcommand{\bq}{\begin{eqnarray}}
\newcommand{\eq}{\end{eqnarray}}
\newcommand{\bsq}{\begin{subequations}}
\newcommand{\esq}{\end{subequations}}
\newcommand{\bc}{\begin{center}}
\newcommand{\ec}{\end{center}}

\title{Analytic Models for the Evolution of Semilocal String Networks}
\author{A.S. Nunes}
\email[Electronic address: ]{c0808014@alunos.fc.up.pt}
\affiliation{Centro de Astrof\'{\i}sica, Universidade do Porto, Rua das Estrelas, 4150-762 Porto, Portugal}
\author{A. Avgoustidis}
\email[Electronic address: ]{A.Avgoustidis@damtp.cam.ac.uk}
\affiliation{C.T.C., DAMTP, University of Cambridge, Wilberforce Road, Cambridge CB3 0WA, U.K.}
\author{C.J.A.P. Martins}
\email[Electronic address: ]{Carlos.Martins@astro.up.pt}
\affiliation{Centro de Astrof\'{\i}sica, Universidade do Porto, Rua das Estrelas, 4150-762 Porto, Portugal}
\author{J. Urrestilla}
\email[Electronic address: ]{jon.urrestilla@ehu.es}
\affiliation{Department of Theoretical Physics, University of the Basque Country UPV-EHU, 48040 Bilbao, Spain}
\affiliation{Department of Physics \& Astronomy, University of Sussex, Brighton, BN1 9QH, United Kingdom}
\date{10 August 2011}
\begin{abstract}
We revisit previously developed analytic models for defect evolution and adapt them appropriately for the study of semilocal string networks. We thus confirm the expectation (based on numerical simulations) that linear scaling evolution is the attractor solution for a broad range of model parameters. We discuss in detail the evolution of individual semilocal segments, focusing on the phenomenology of segment growth, and also provide a preliminary comparison with existing numerical simulations.
\end{abstract}
\maketitle

\section{Introduction}

Cosmic strings are line-like concentrations of energy that can arise as topological defects in cosmological phase transitions in the early universe \cite{Kibble80,VilShell}. Their notable popularity in the 80's was due to the belief that they could be responsible for structure formation, an alternative scenario to inflationary perturbations \cite{Hawking82,Mukhanov82,Linde82}.  At the end of the 90's, however, it was realised \cite{AlbrBat,AlbrBat1} that strings cannot be solely responsible for the observed structure, which was further confirmed by the direct observations of acoustic peaks in the CMB, in accordance with the inflationary prediction. This led to a shift of attention away from cosmic strings.  

However, recent developments in fundamental theory have led to a resurgence of interest in the subject \cite{KibbleRev,PolchIntro}.  First, it was realised that the production of cosmic strings is much more generic than previously thought in a wide class of cosmological models based on supergravity 
\cite{Jeannerot}.  In addition, new exciting developments in string phenomenology 
\cite{PolchBranes,KiritsisRev,QuevedoRev} allowed the study of cosmological inflation 
from brane constructions, and this led to a generic picture of hybrid inflation ending in a phase transition that produces cosmic (super)strings \cite{BMNQRZ,MajumDavis,DvalTye}.

These objects are much more complex than ordinary field theory strings and their properties depend sensitively on fundamental high-energy physics parameters of the underlying theory  
\cite{JoStoTye,PolchProb,PolchStab}. Thus, not only it was realised that string-like defects are much more generic objects that could play a subdominant role in structure formation, but, also, their potential observability could open a window into physics at the highest energy scales.          

Over the past few years, there has been significant progress in understanding the properties, cosmological evolution and possible observational signatures of cosmic superstrings (for a recent review see \cite{CopPogVach}).  However, there remain certain types of well-motivated string objects for which their physics, and most notably their cosmological evolution, remains relatively unexplored. For example, there is yet no detailed self-consistent model describing the evolution of cosmic strings carrying currents, even though such strings have been predicted in situations where more than one cosmological phase transition is considered \cite{Brandenberger:1996zp}, and they also arise naturally in SUSY models that predict strings  \cite{Davis:1997bs,Achucarro:2002jg}. String currents also appear in the effective 4D description of higher dimensional strings, as the string position in the internal dimensions is described by worldsheet scalar fields giving rise to currents.  Another example is semilocal strings \cite{VachAchuc}, arising in SUSY GUT inflationary models \cite{Urrestilla:2004eh} and the corresponding D3/D7 brane inflation models \cite{D3D7semiloc}. These models are a natural extension of usual inflationary models, in which the only extra ingredient needed is doubling a hypermultiplet. This simple change renders the properties of the strings very different; as we will discuss in the following section, these semilocal strings are non-topological, they have different stability properties ~\cite{Vachaspati:1991dz,Hindmarsh:1991jq} and even their CMB constraints are less stringent than those for usual cosmic strings~\cite{Urrestilla:2007sf}. 

More recently, monopole networks and hybrid networks (where finite string segments have monopoles attached to their ends) have also been studied \cite{MartAch,Hybrid}. Both in the gauge and in the global case, the hybrid networks tend to annihilate very quickly, usually in less than a Hubble time after formation, although they may survive for a longer period in particular circumstances. There is an outstanding hybrid-like case, where local strings attached to global monopoles. This, at least naively, is the starting point for the much more interesting semilocal case which we discuss here. 

String evolution is a difficult problem involving physics from different energy  scales, which is often not fully understood. There have been many different approaches for obtaining the properties of string networks, trying to negotiate between computational power and accuracy, between numerical and analytic approaches. Numerical simulations approaches include field theoretic simulations, Nambu-Goto approximations, and the so-called unconnected segment toy models \cite{AlbrBat}. All these approaches have their virtues and their limitations, and progress has been possible through combination  of results from different fronts. Another complementary approach is that of analytic velocity-dependent one-scale (VOS) models \cite{VOS96,VOS02}. These abandon a `statistical physics' description of the network in favour of a `theromodynamics' one, characterising the network by a small number of physically meaningful macroscopic quantities such as correlation lengths and root-mean-squared (RMS) velocities, whose evolution equations can be calculated from the microscopic equations. This approach has the obvious advantages of simplicity, tractability and applicability beyond the dynamical ranges accessible to simulations, but it comes at a cost---the averaging process (through which one deduces the macroscopic evolution equations from the microscopic ones) requires the introduction of a (relatively small) number of phenomenological parameters, whose values can only be inferred by using numerical simulations for calibration.

In what follows we discuss such an analytic model for the evolution of semilocal string networks. We study the behaviour of the network as a whole, starting from the premise that it can be treated as a network of local strings attached to global monopoles. We also describe the evolution of individual semilocal segments, discussing under what conditions these segments can grow---a process that has been clearly identified in numerical simulations. Finally, we provide a preliminary comparison between the analytic model and existing numerical simulations, focusing on the overall network properties and the
dependence on the coupling parameter. A more detailed comparison, discussing in detail the distribution of the semilocal segments and its time evolution, is left for a follow-up paper.

\section{Semilocal Strings}
\label{semi}

Semilocal strings  \cite{Achucarro:1999it,Vachaspati:1991dz,Hindmarsh:1991jq}  were introduced as a minimal extension of the Abelian Higgs model with two complex scalar fields instead of one, which lead to $U(1)$ flux-tube solutions even when the vacuum manifold is simply connected. The strings of this extended model will share properties with ordinary local $U(1)$ strings, but they are not purely topological and will have different properties. For example, since they are not topological, they need not be infinite and can have ends. These ends are effectively global monopoles, which have some exotic properties by themselves~\cite{Achucarro:2000td}. 

It is thus clear that this semilocal string model is a field theoretical version of the type of object that we want to study in this paper (strings attached to monopoles). In section~\ref{comp}, we will use numerical simulations performed using this field theoretical model  to verify the results obtained in this paper, so we provide the reader with a short introduction to the semilocal string model in this section.

The relevant action for the simplest semilocal string model, the one we will use in the numerical simulations of section~\ref{comp}, reads
\be
S=\int\!\! d^4x\!\left[\left[(\partial_\mu-i A_\mu)\Phi\right]^2-\frac{1}{4}F^2-\frac{\beta}{2}(\Phi^+\Phi-1)^2\right]
\label{SLaction}
\ee
where $\Phi=(\phi,\psi)$. It can be easily seen that setting one of the two scalar fields to zero, we recover the Abelian Higgs model. We can therefore build from the analytical models applied to usual cosmic strings  to tackle this new problem. 

The stability of the strings is not trivial, and it will depend on the value of the parameter $\beta=m^2_{\rm scalar}/m^2_{\rm gauge}$:  for $\beta<1$ the string is stable, for $\beta>1$ it is unstable, and for $\beta=1$ it is neutrally stable. As we will see in section~\ref{comp}, only low values of $\beta$ will be of interest for the comparison, because otherwise the string network is either unstable or disappears very fast~\cite{Achucarro:2007sp}.

After a cosmological phase transition in such a model, it is expected that segments of semilocal strings will form. The cosmological evolution of a semilocal segment network will be quite different from the evolution of ordinary Abelian-Higgs strings~\cite{Benson:1993at,Achucarro:1999it}. The fact that semilocal strings have a different cosmological evolution is interesting because CMB predictions can be different~\cite{Urrestilla:2007sf} and can aid on inflationary model building~\cite{Urrestilla:2004eh}.

The network evolution will depend on the interplay between string dynamics and monopole dynamics.  When the string segment ends, it must end in a cloud of gradient energy. Those string ends will behave like global monopoles that  provide an interaction between strings independent of distance. Therefore, depending on the interplay between string dynamics and monopole dynamics, the segments can contract and eventually disappear, or they can grow to join a nearby segment and form a very long string~\cite{Achucarro:2007sp}. 

We thus see that, at least to a first approximation, we can envisage these networks as being made of local strings attached to global monopoles, and as such the previously developed analytic modelling  techniques~\cite{VOS96,MartAch} should be applicable. This being said, it is also clear that these networks possess non-trivial dynamical properties, beyond those of standard hybrid networks \cite{VOS96,MartAch,Hybrid}. Specifically, the evolution of the string network will depend both on the string tension and on the dynamics of the gradient energy: the latter may be thought of as providing a long-range interaction between the strings. Nevertheless, we can hope to factor this in by modifying and extending these models so as to include the relevant effects. 

\section{Monopole Network Evolution}

A simple way to treat semilocal strings is by analogy with the standard case of monopoles connected by strings, the difference being that the strings are gauged but the monopoles are global. We emphasise that in the present section we will describe the network by explicitly modelling the dynamics and interactions of the monopoles. (In the following section we will follow the complementary approach and look at the evolution of individual string segments.) This is particularly justified in the semilocal case since (as has been shown in previous work \cite{Achucarro:2007sp}) it is indeed the monopoles that control the evolution of the network.

The standard approach to analytical modelling of defect networks is to start from the microscopic 
equations of motion -- e.g. the Nambu-Goto equations of motion for string defects -- and perform 
a statistical averaging procedure, under the assumption that the defects are randomly distributed 
at large enough scales.  This allows one to construct a macroscopic energy evolution equation 
(which can be traded for an equation for the network's characteristic lengthscale) and an equation 
for the RMS network velocities, which together describe the network at 
large-scales in a 'thermodynamical' sense.  Defect interactions are then added to these equations 
in a phenomenological way.  In the case of cosmic strings, this procedure leads to the so-called 
VOS model \cite{VOS96,VOS02}, which has been well-tested 
against simulations and is used for predicting CMB signals of string networks.   
    
In the case of monopoles, one can follow an analogous procedure.  The idea is to write an 
evolution equation for the monopole density, neglecting interactions, and then re-express it
in terms of  a characteristic lengthscale, $L$  (the average inter-monopole distance).  The 
effects of monopole forces and friction are then included in this equation (as well as in the
relevant velocity equation) by adding extra phenomenological terms.  More details can be 
found in \cite{MartAch,Hybrid}. The evolution equation for the characteristic monopole lengthscale 
reads as follows
\be
3\frac{dL}{dt}=3HL+v^2\frac{L}{\ell_d}+c_\star v\,,
\ee
where $c_\star$ is a free parameter (to be calibrated by simulations) quantifying energy loss, and where 
we have defined a damping lengthscale, $l_d$ that includes the effects of expansion (Hubble parameter $H$) 
and of friction (lengthscale $l_f$) due to particle scattering
\be
\frac{1}{l_{d}}=H+\frac{1}{l_{f}}\,.
\end{equation}
The evolution equation for the RMS velocity $v$ of the monopoles is
\be
\frac{dv}{dt}=(1-v^2)\left[\frac{k_m}{L}\left(\frac{L}{d_H}\right)^{3/2}+\frac{k_s}{L}\frac{\eta_s^2}{\eta_m^2}-\frac{v}{\ell_d}\right]\,,
\ee
where the second term in the square bracket describes the force due to the strings, and for an expansion rate of the generic form $a(t)\propto t^{\lambda}$, the Hubble parameter and Hubble distance are respectively 
\be
H=\frac{\lambda}{t}\,,\quad d_{H}=\frac{t}{1-\lambda}\,.
\ee
The constants $k_m$ and $k_s$ parameterise the monopole and string forces, and $\eta_s$, $\eta_m$ 
are the relevant symmetry breaking scales.  Since we are mostly interested in late-time scaling solutions we will (unless otherwise stated) neglect the effect of friction due to particle scattering, which is only relevant in the early stages of the network's evolution.

For most of this and the following section we shall consider standard expansion rates, corresponding to the parameter range $0<\lambda <1$, so that $\lambda=1/2$ in the radiation-dominated era and $\lambda=2/3$ in the matter-dominated era. (Towards the end of the current section we will briefly comment on the limiting cases $\lambda=0$ and $\lambda=1$.) The reason for explicitly studying radiation and matter era solutions is that recent observational constraints \cite{Urrestilla:2007sf} show that semilocal string networks cannot be the dominant component of the universe's energy budget, but can only contribute a small fraction to it.

Immediately we see a difference relative to the other hybrid networks. In \cite{Hybrid} it was shown that for local strings attached to local monopoles the force due to the strings was always dominant when compared to that due to the monopoles, while for global strings attached to global monopoles the string force was subdominant at string formation but became dominant later in the network's evolution. In the present case the ratio is
\be
\frac{f_s}{f_m}=\frac{k_s}{k_m}\left(\frac{\eta_s}{\eta_m}\right)^2\left(\frac{d_H}{L}\right)^{3/2}
\ee
and therefore the string force should always be subdominant. This is in agreement with theoretical expectations and numerical simulations.

Interestingly, this combination of terms means that the only attractor solution in an expanding universe (with $a\propto t^\lambda$) is linear scaling
\be
L=\gamma t\,,\quad v=v_0\,,\label{linscal}
\ee
as in the case of plain global monopoles, and indeed we can closely follow the discussion in \cite{MartAch}. In particular, there are two possible branches of the scaling solution. First, there is an ultrarelativistic one with
\be
\gamma=\frac{c_\star}{3-4\lambda}\,,\quad v_0=1\,,
\ee
which only exists for relatively slow expansion rates ($\lambda<3/4$) but is in principle allowed both on the radiation and matter eras. Second, a normal solution exists for any expansion rate, and its scaling parameters obey
\be
\gamma = \frac{c_\star v_0}{3-\lambda(3+v_{0}^{2})} \,, \label{nscaling1}
\ee
\be
\lambda v_0=k_m(1-\lambda)^{3/2}\gamma^{1/2}\,. \label{nscaling2}
\ee
As in the case of standard global monopoles there is a constraint on the velocities
\be
v_0^2<3(\frac{1}{\lambda}-1);
\ee
which is trivial for $\lambda<3/4$ (that is, $v_0\to1$ is allowed), but restrictive for faster expansion rates. On the other hand, velocities will generically be significant: having $v_0\to0$ requires $\lambda\to1$. 

\begin{figure}
\includegraphics[width=3.3in,keepaspectratio]{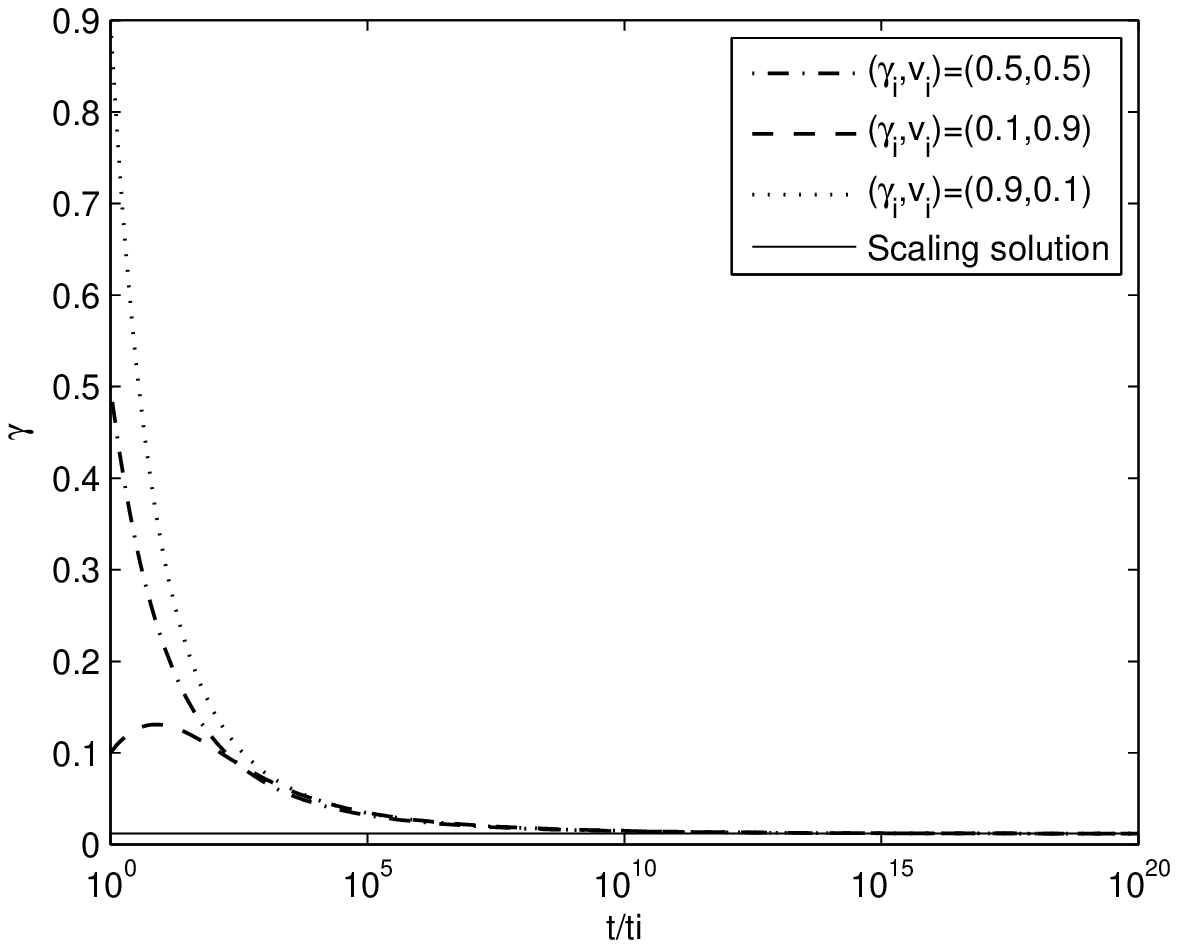}
\includegraphics[width=3.3in,keepaspectratio]{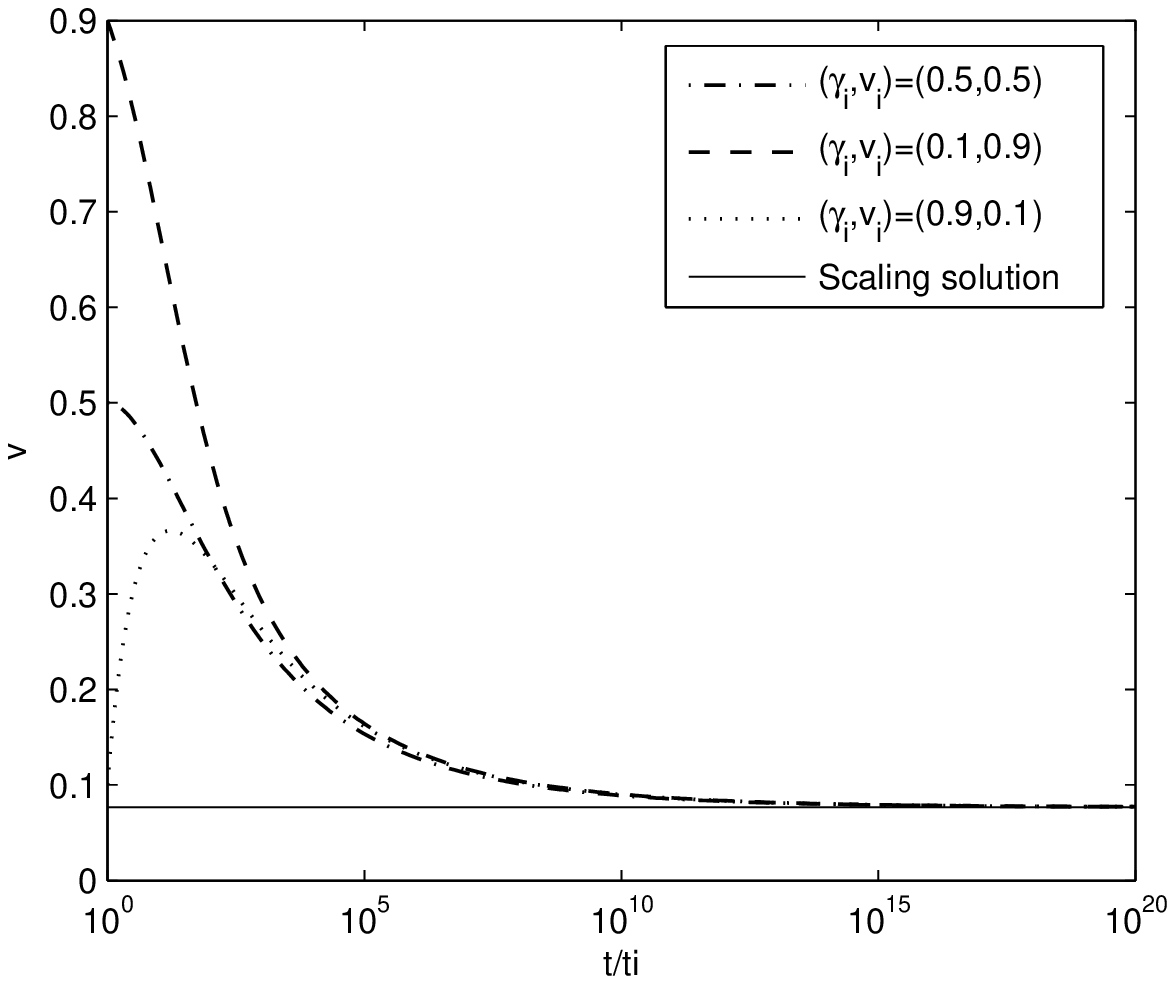}
\caption{\label{fig1}The time evolution of $\gamma$ (top panel) and $v$ (bottom panel), for three different choices of initial conditions, with the parameter choices $\lambda=1/2$, $k_{m}$\emph{=}1 and $c_\star=0.23$.}
\end{figure}

In what follows we briefly explore the parameter space of these scaling solutions. In Fig. \ref{fig1} we show an example of the convergence of $\gamma$ and $v$ towards the scaling solution in the radiation era, which for this particular choice of parameters is $\gamma_{o}=0.0118$, $v_{o}=0.0768$.

\begin{figure}
\includegraphics[width=3.3in,keepaspectratio]{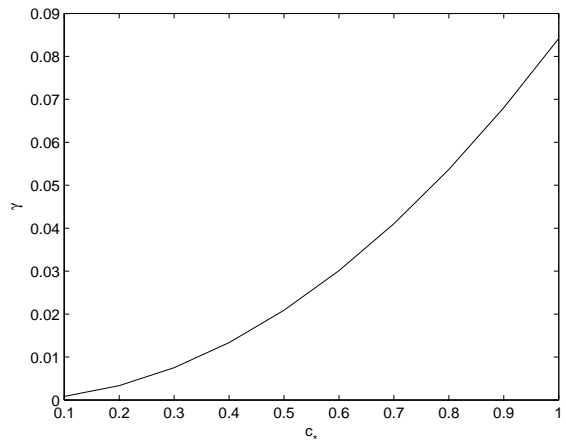}
\caption{\label{fig2}The dependence of the scaling parameter $\gamma$ on $c_\star$, for $\lambda=2/3$ and $k_{m}$\emph{=}1.}
\end{figure}

In Fig.~\ref{fig2} we illustrate the dependence of $\gamma$ on $c_\star$, with other parameters kept fixed: the slow growth is noteworthy. Since strings typically have lower energies than the monopoles, the energy loss parameter $c_\star$ mostly reflects the annihilation of monopoles and antimonopoles, with some contribution from radiative losses; loop formation losses are expected to be much smaller. Therefore increasing $c_\star$ corresponds to increasing the annihilation rate, and the typical monopole separation correspondingly increases.

\begin{figure}
\includegraphics[width=3.3in,keepaspectratio]{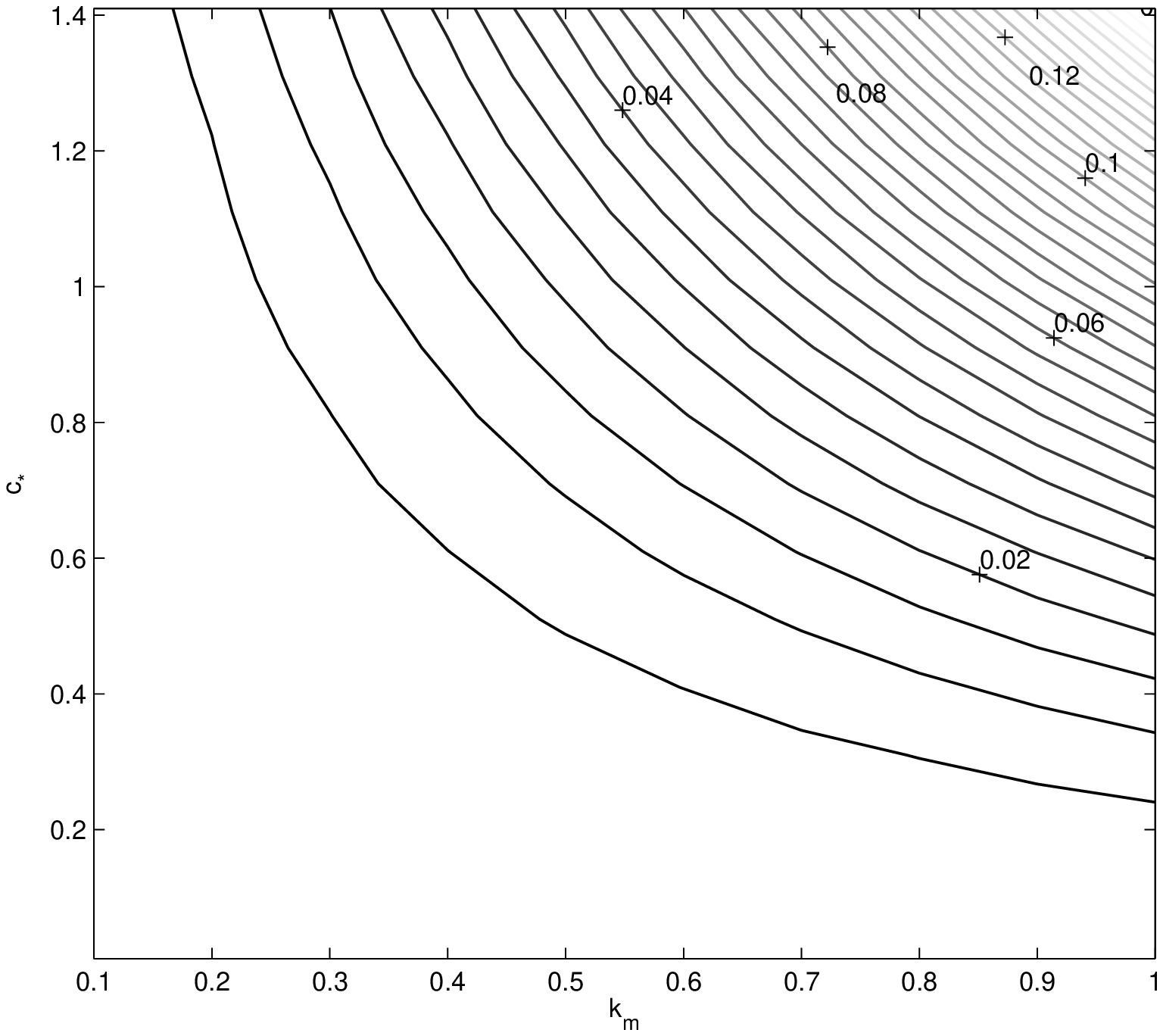}
\includegraphics[width=3.3in,keepaspectratio]{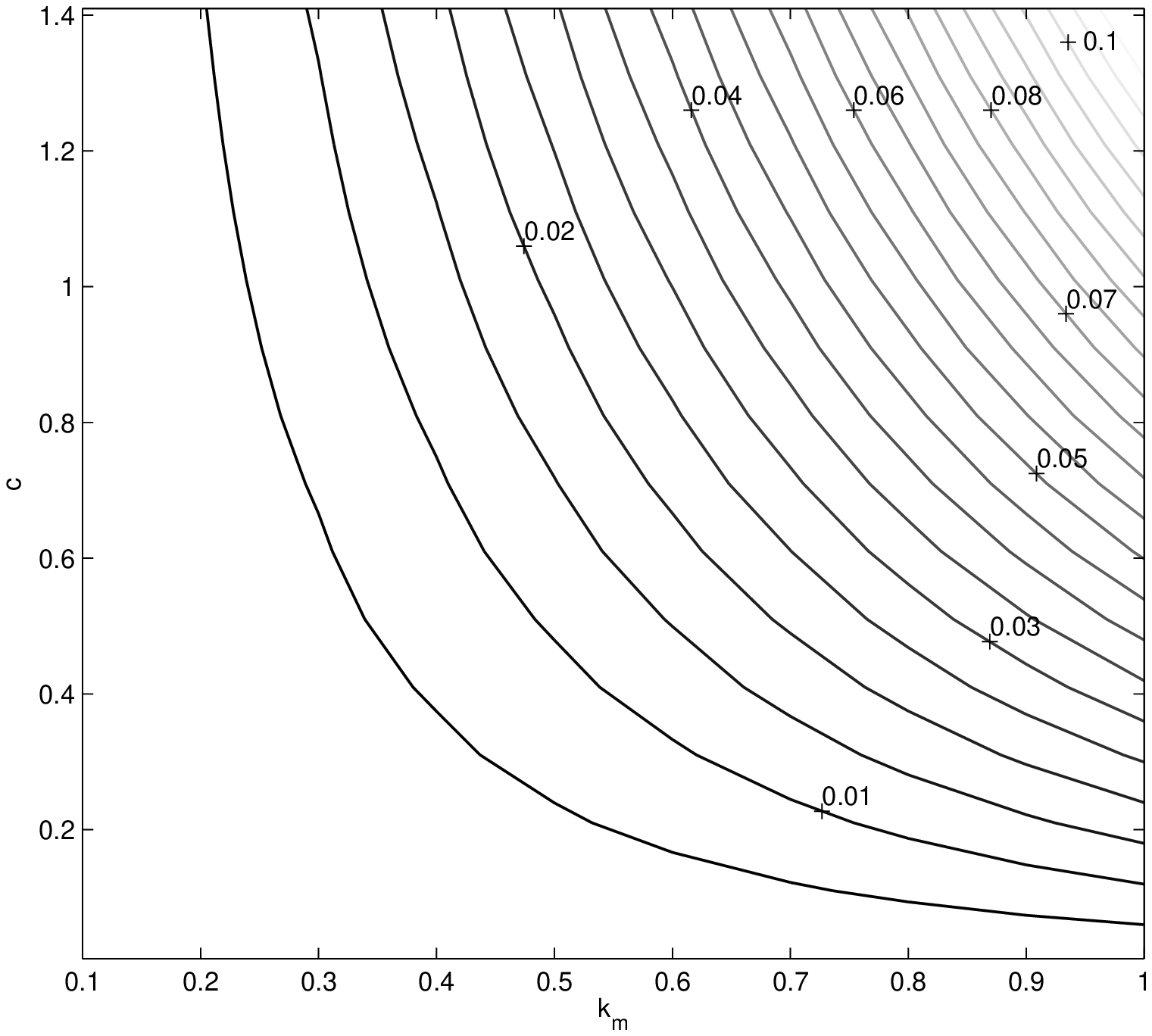}
\caption{\label{fig3}The value of the asymptotic scaling parameters $\gamma$ (top panel) and $v$ (bottom panel) as a function of $c_\star$ and $k_{m}$, for the matter-dominated era, $\lambda=2/3$.}
\end{figure}

Figure~\ref{fig3} generalises the above and displays the dependence of the asymptotic scaling values of $\gamma$ and $v$ on the two parameters $c_\star$ and $k_{m}$, in this case for the matter epoch. As expected $\gamma$ also grows slowly with $k_{m}$, and $v$ grows with both parameters. 

\begin{figure}
\includegraphics[width=3.3in,keepaspectratio]{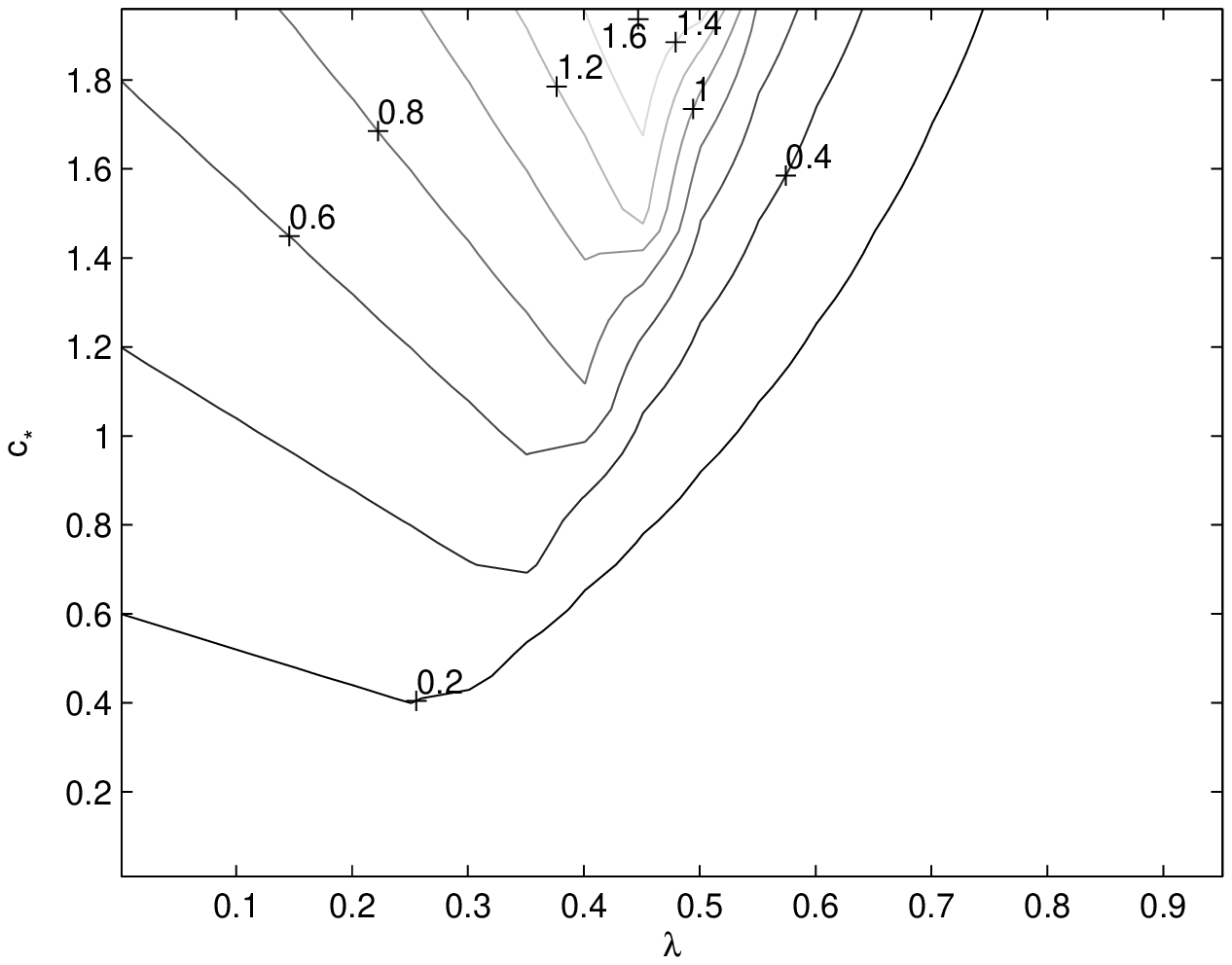}
\includegraphics[width=3.3in,keepaspectratio]{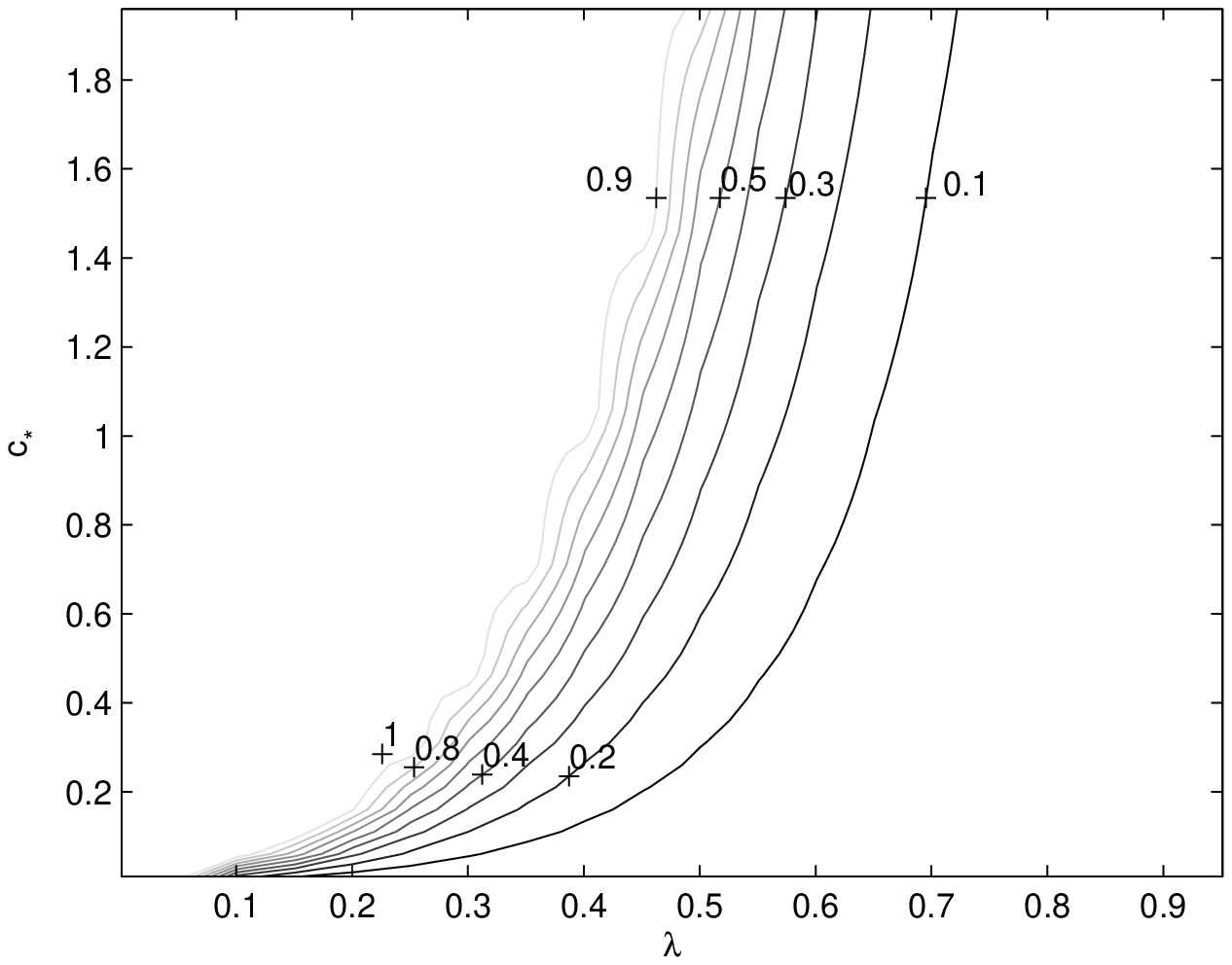}
\caption{\label{fig4}The value of the asymptotic scaling parameters $\gamma$ (top panel) and $v$ (bottom panel) as a function of $c_\star$ and the expansion rate parameter $\lambda$, for a fixed $k_m=1$.}
\end{figure}

Finally, Fig.~\ref{fig4} quantifies the effect of the expansion rate, by displaying the asymptotic values of $\gamma$ and $v$ as a function of $c_\star$ and $\lambda$, for a fixed $k_m=1$. Here the behaviour is more interesting. One sees that $\gamma$ has rapid and non-trivial variations for slow expansion rates, and inspection of the $v$ plot reveals that this corresponds to the network being in the ultrarelativistic branch, and eventually (at moderate expansion rates) switching to the normal one.

We note that it is also of interest to consider the case of Minkowski space (corresponding to $\lambda=0$ and $H=0$) but with a friction lengthscale proportional to the correlation length (say, for simplicity, $\ell_f\sim L$). This should be an adequate description of some of the numerical simulations of semilocal strings done so far \cite{Achucarro:2007sp}. In this case, linear scaling is still the attractor solution, although the scaling parameters now obey different relations, namely
\be
3\gamma =v_0^2+cv_0\,,\quad v_0=k_m\gamma^{3/2}\,.
\ee
It might be possible to test these relations numerically, in particular by comparing simulations done with different values of the parameter $\beta$.

On the other hand, in the opposite limit of fast expansion rate ($\lambda\ge1$, or in other words inflation) the linear scaling solution of Eq. \ref{linscal} no longer exists. In this case the network is conformally stretched and approximately frozen, and its scaling solution is
\be
L\propto a\,,\quad v\propto \frac{1}{HL}\,.
\ee
These conformal stretching solutions are ubiquitous in the defects literature.

The analysis in this section may be simplistic in several ways, although the result of linear scaling is encouraging, since it is seen in numerical simulations. The key issue that still needs to be considered is the role and consequences of the gradient energy. It may be thought of as providing a long-range interaction between the strings, but it is not clear what its dynamical effect on the monopoles will be.

A possible way for describing the effects of the gradient energy is to treat semilocal networks as composed of two types of strings (standard strings and gradient strings). In this description an important issue is what happens at the string ends (monopole junctions)---there will be some matching condition that would need to be encoded in the evolution equations.

\section{Semilocal Segment Evolution}

\begin{figure}
\includegraphics[width=3.3in,keepaspectratio]{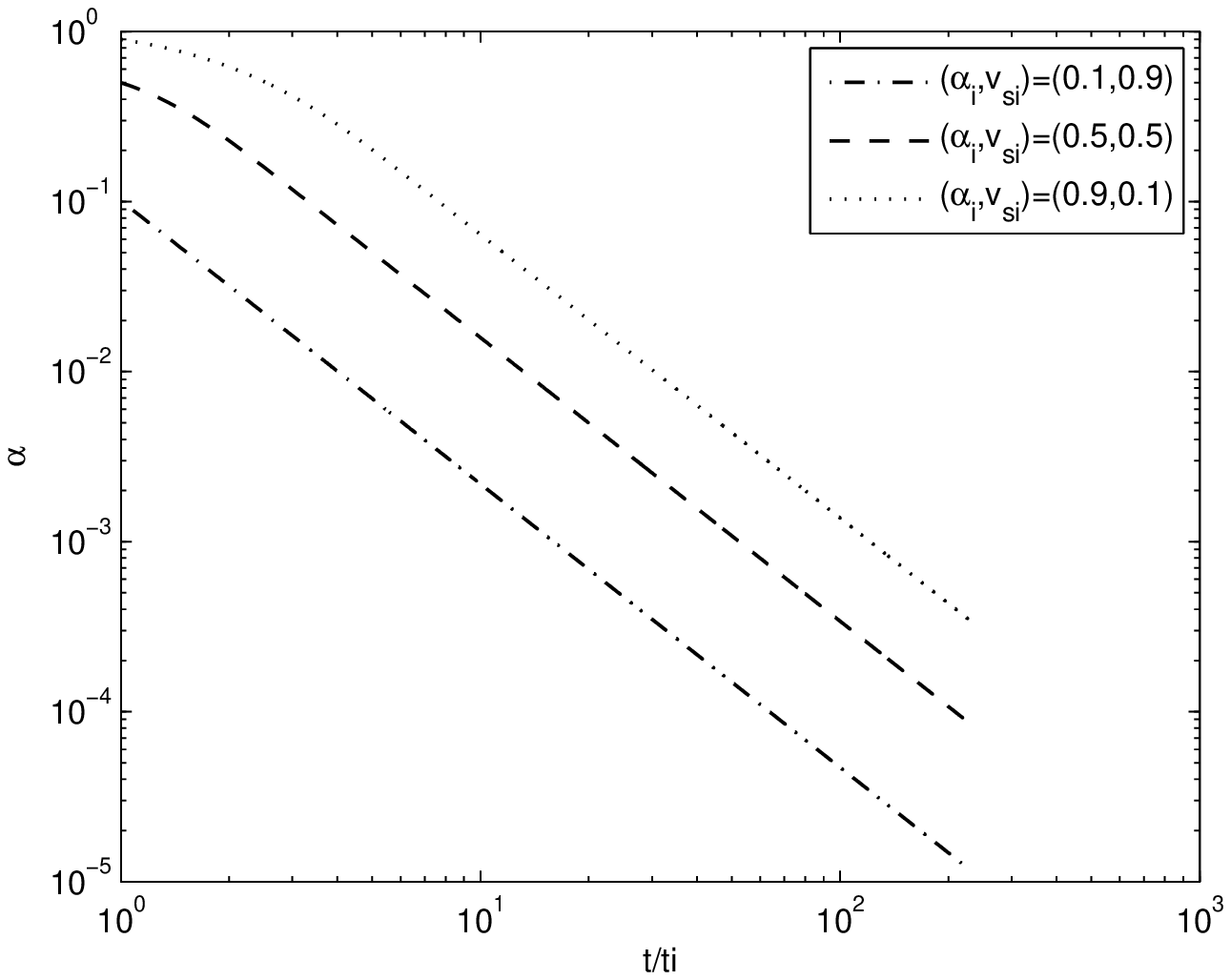}
\includegraphics[width=3.3in,keepaspectratio]{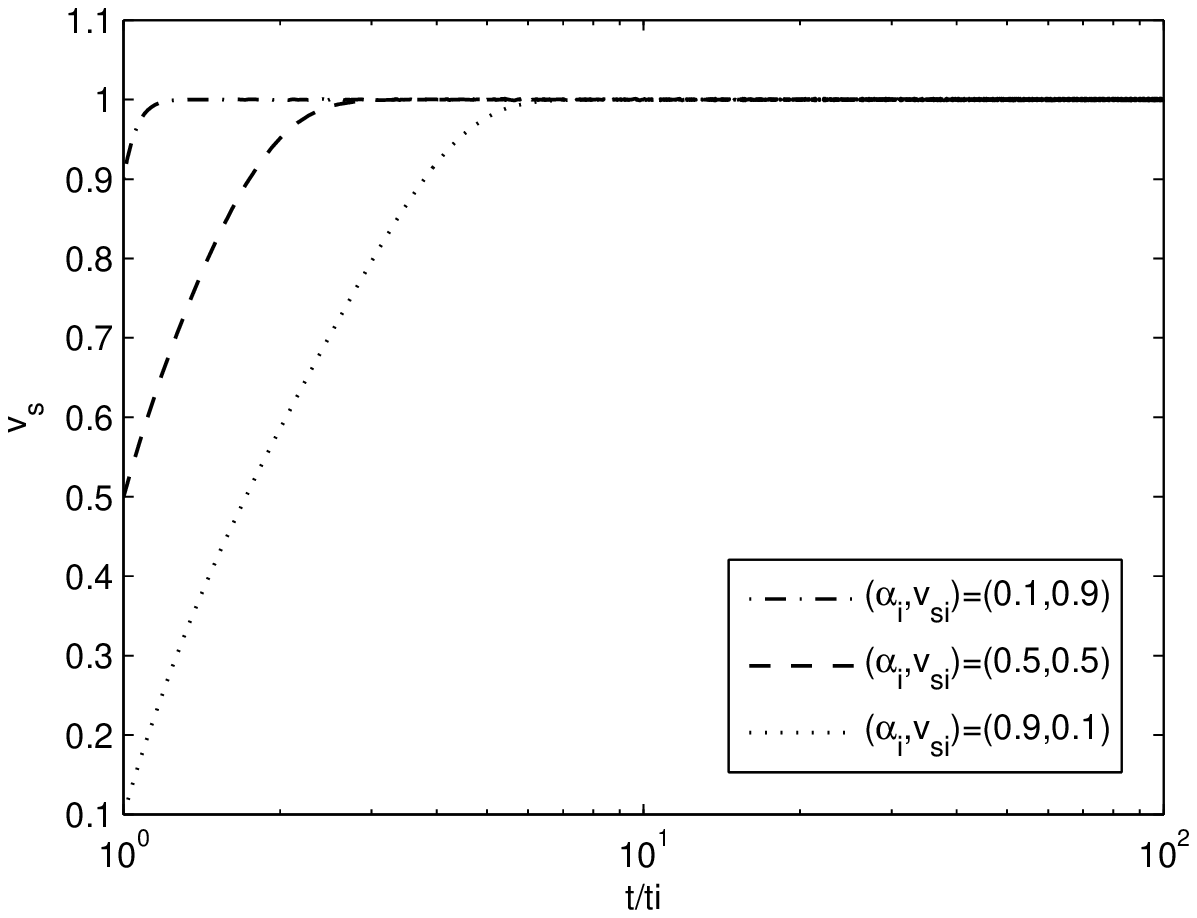}
\caption{\label{fig5}The time evolution of $\alpha=l_{s}/t$ (top panel) and $v_s$ (bottom panel) for string segments with three different initial conditions, for a matter-dominated era and $k=1$.}
\end{figure}

Even if we can characterise the network by focusing on the evolution of the monopoles, it is still relevant to understand the evolution of the individual string segments, although this may be possible only in an averaged sense. The evolution equations for string segments can straightforwardly be derived in the context of the VOS model \cite{VOS96}, yielding
\be
\frac{dl_{s}}{dt}=Hl_{s}-v_{s}^{2}\frac{l_{s}}{l_{d}}
\ee
\be
\frac{dv_{s}}{dt}=\left(1-v_{s}^{2}\right)\left[\frac{k}{l_{s}}-\frac{v_{s}}{l_{d}}\right]\,,
\ee
where $l_{s}$ is the length of the segment under consideration, $v_{s}$ its RMS velocity, $k$ a 
parameter describing string curvature, and $l_{d}$ is now the string damping length,
\be
\frac{1}{l_{d}}=2H+\frac{1}{l_{f}}\,.
\ee
Again, we will neglect the damping effect of friction from particle scattering, and consider only that due to the Hubble expansion. 
At this point we are also neglecting energy losses in the segments due to radiation or loop formation; we will come back to these issues shortly. Naturally, in these circumstances all segments will quickly shrink, become ultrarelativistic and disappear, as illustrated in Fig. \ref{fig5}: the segment size relative to the horizon, $\alpha=l_{s}/t$, and the segment velocity respectively follow $\alpha\rightarrow0$ and $v_s\rightarrow1$.

However, theoretical expectations, confirmed by numerical simulations, indicate that the dynamics of semilocal string segments is more interesting than that. While some segments do disappear, there is also a significant probability that segments merge and form longer segments. This is due to the long range interactions of the global monopoles at the end of the segments, as has been discussed in Sect. II. In certain circumstances this process could even culminate in the disappearance of the monopoles, leaving behind a network of standard Abrikosov-Nielsen-Olesen strings.

In what follows we discuss two possible phenomenological ways of accounting for this behaviour, by modifying the evolution equation for the segment size. A possible alternative  would be to modify the forces in the velocity equation, keeping in mind the notion that at large distances the force should be string-like (inversely proportional  to distance) but at short distances it may be independent of distance. This alternative is left for subsequent work.

\subsection{Scale-dependent Behaviour}

A simple generalisation of the above equations would be
\be
\frac{dl_{s}}{dt}=Hl_{s}-v_{s}^{2}\frac{l_{s}}{l_{d}}+\sigma\left(1-\frac{L}{l_{s}}\right)v_{m}^{2} \label{dlsdt}
\ee
\be
\frac{dv_{s}}{dt}=\left(1-v_{s}^{2}\right)\left[\frac{k}{l_{s}}+f_{s}-\frac{v_{s}}{l_{d}}\right],
\ee
where $L$ is the characteristic scale of the monopoles (and $v_m$ their characteristic velocity) that has been previously discussed, and $\sigma$ is a free parameter controlling the importance of the newly introduced term. We can in principle allow for a force due to the strings in the velocity equation, but this is expected to be subdominant, so we will neglect it in what follows; a further simplifying assumption (justified from our analysis in the previous section) is to take $v_m\sim1$ 
in Eq.~(\ref{dlsdt}).

The new term was added on the purely phenomenological reasoning that, to a first approximation, small segments should shrink and large ones should grow and merge. This can be intuited as a competition between two characteristic timescales. Each segment will have an annihilation timescale, and each monopole will have a characteristic timescale in which to find its (anti)partner, annihilate, and therefore produce a longer segment. The second process is expected to become relatively more likely as the segment size increases.

One can search for scaling solutions of these equations, of the form $\alpha=l_{s}/t=const.$, $v_{s}=const.$. Here we will assume that $v_m=1$ and also that the overall density of the network has itself reached scaling, so that $L=\gamma t$ with $\gamma=const.$ as previously discussed. Again one finds two possible branches. There is an ultrarelativistic branch with
\be
v_{s}=1\,,\quad \alpha^{2}=\sigma\frac{\alpha-\gamma}{\lambda+1},
\end{equation}
and also a normal one with
\be
\alpha^{2}=\sigma\frac{\alpha-\gamma}{\lambda\left(2v_{s}^{2}-1\right)+1}\label{quadratic1}
\ee
\be
v_{s}=\frac{k}{2\lambda\alpha}\,.
\ee
Defining $\Gamma=\lambda\left(2v_{s}^{2}-1\right)+1$, Eq.~(\ref{quadratic1}) has two real solutions, provided
\be
\sigma>4\Gamma\gamma\,;\label{cond1}
\ee
if this is the case, then
\be
\alpha=\frac{\sigma}{2\Gamma}\pm\sqrt{\left(\frac{\sigma}{2\Gamma}\right)^{2}-\frac{\sigma}{\Gamma}\gamma}\,.
\ee
Note that the negative sign solutions correspond to $\alpha\sim\gamma$. If $\sigma$ is too small to fulfil the condition of Eq.~(\ref{cond1}), then the asymptotic behaviour is as in the standard case ($\sigma=0$).

\begin{figure}
\includegraphics[width=3.3in,keepaspectratio]{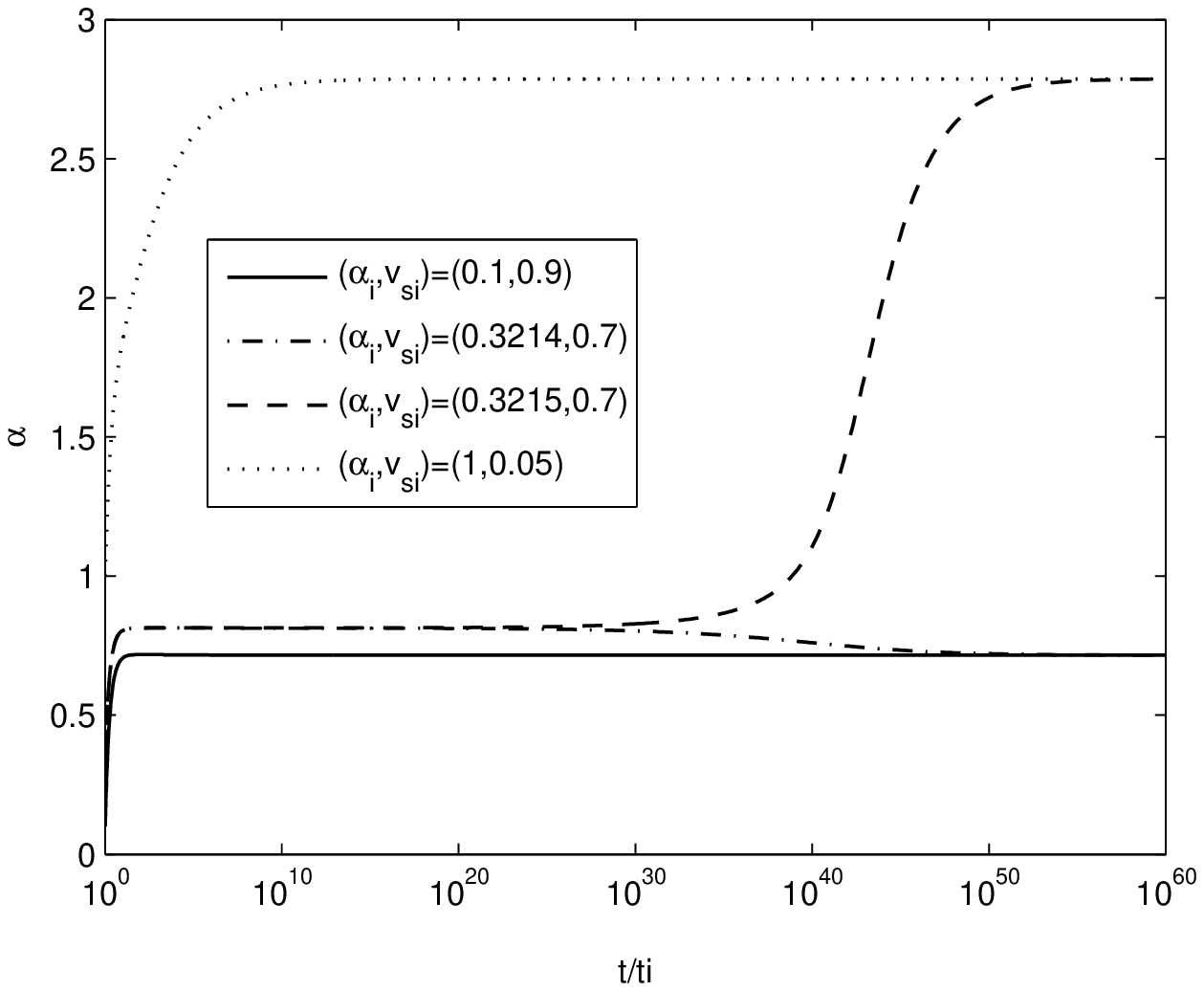}
\includegraphics[width=3.3in,keepaspectratio]{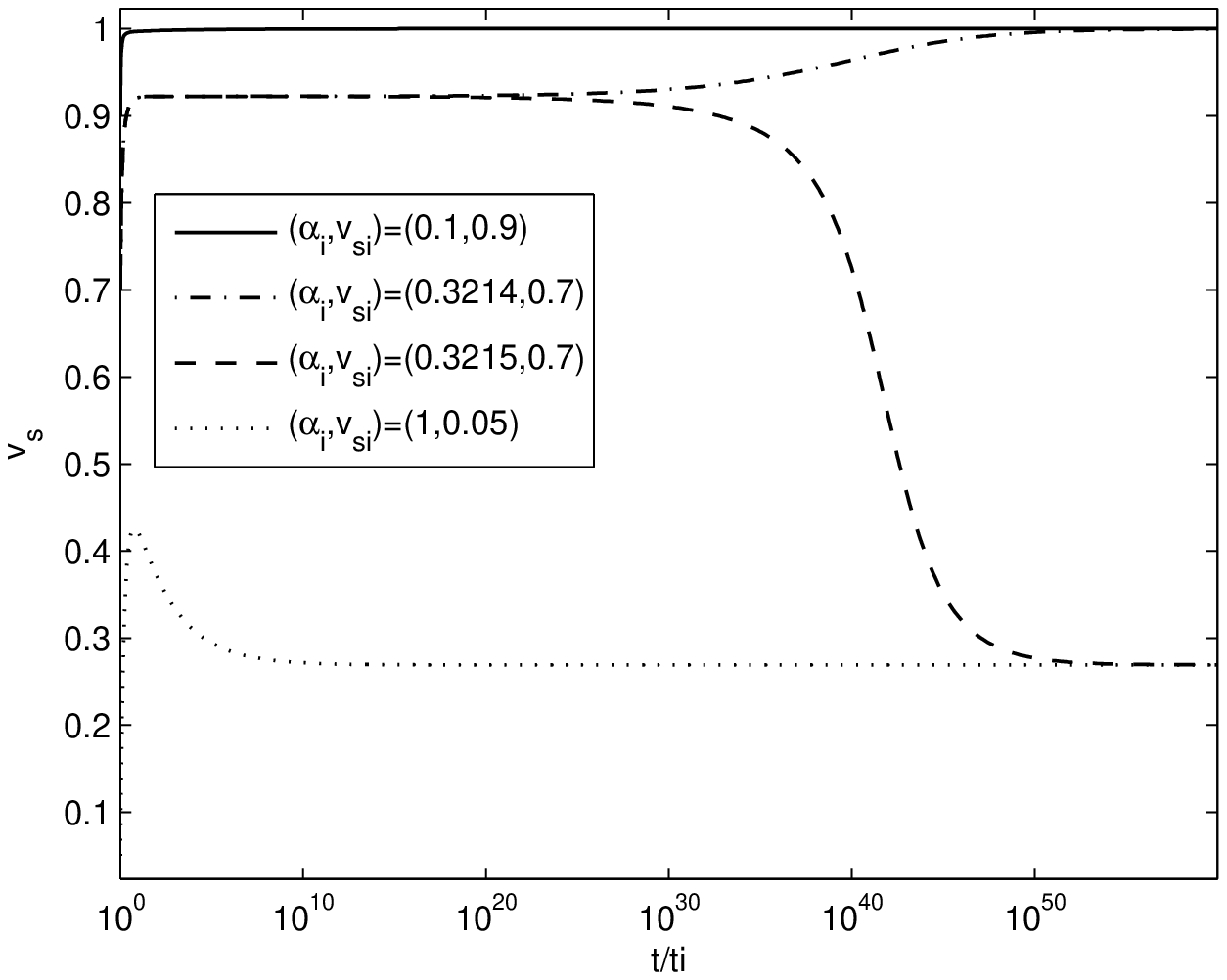}
\caption{\label{fig6}The time evolution of $\alpha=l_{s}/t$ (top panel) and $v_s$ (bottom panel) for string segments with different initial conditions in a matter-dominated era, with the parameter choices $\sigma=1.2$, $k=1$.}
\end{figure}

The above behaviour is illustrated in Fig. \ref{fig6}, for $\sigma=1.2$, a matter dominated era and the parameter choices $c_\star=0.23$ (for the monopole network) and $k=1$. In this case the asymptotic values of $\alpha$ are respectively $\alpha=2.787$ and $\alpha=0.7156$, while the corresponding velocities are $v_{s}=0.2691$ and $v_{s}=1$. For some initial conditions one also observes a transient scaling solution. Here, therefore, the subsequent evolution of a segment, in particular which of the two branches of the scaling solution it follows, will depend on its initial conditions.

\begin{figure}
\includegraphics[width=3.3in,keepaspectratio]{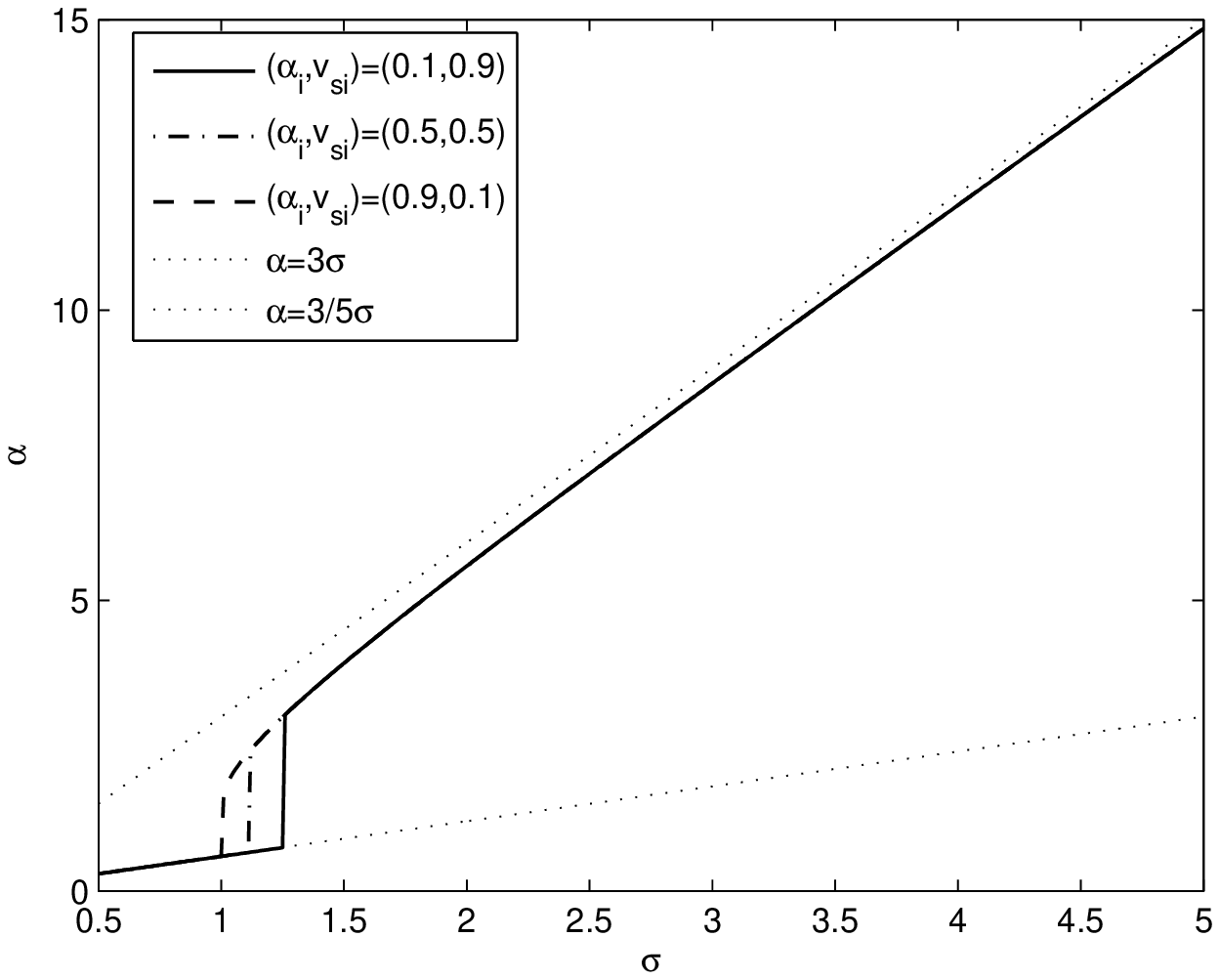}
\includegraphics[width=3.3in,keepaspectratio]{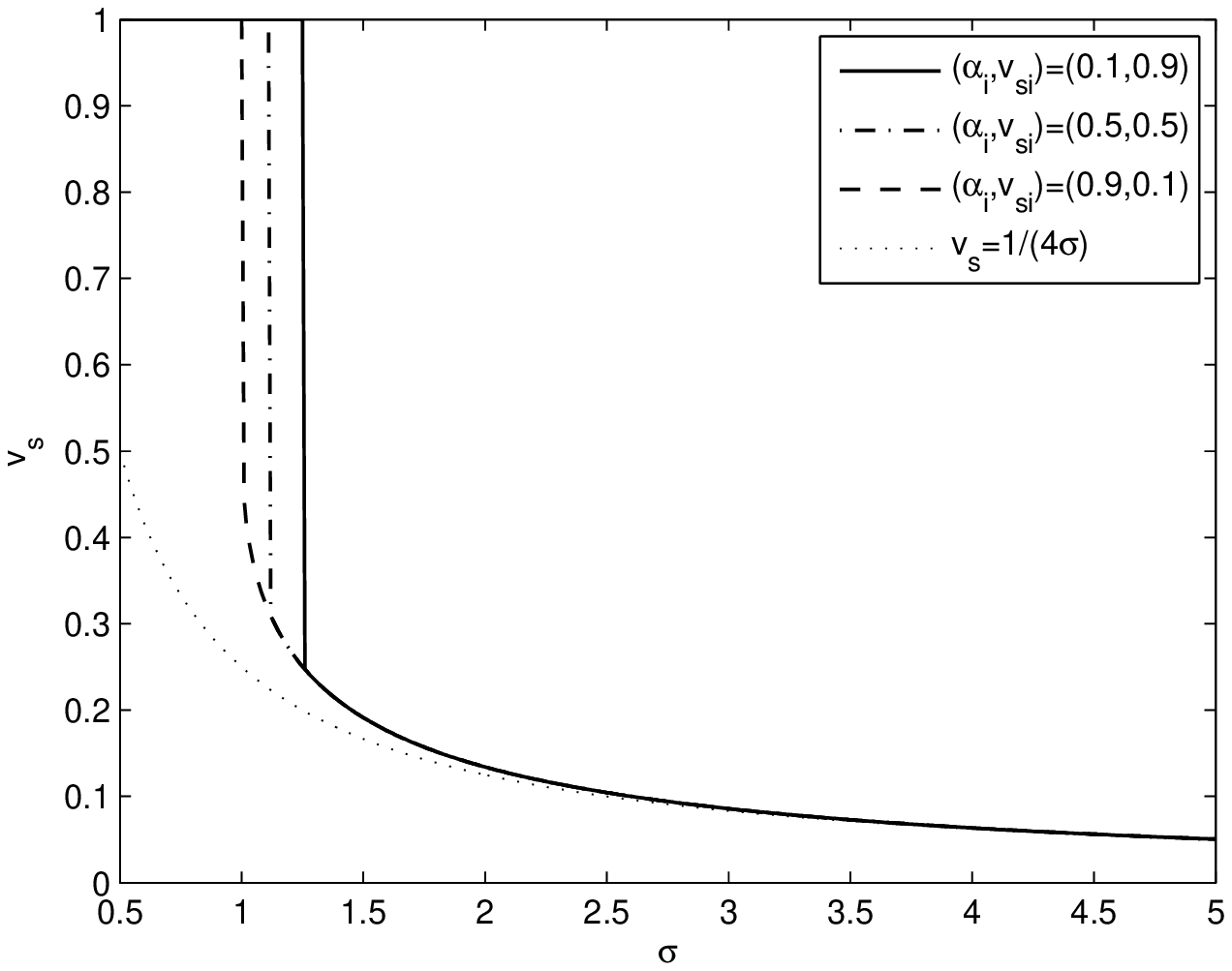}
\caption{\label{fig7}The scaling behaviour of $\alpha$ (top panel) and $v_s$ (bottom panel) as a function of $\sigma$, for segments of different initial sizes and velocities; the plots are for the matter-dominated era, with $c_\star=0.23$ and $k=1$. Also shown with dotted lines are the asymptotic behaviours in the limits of large and small $\sigma$, which are further discussed in the text.}
\end{figure}

One can also determine how the asymptotic scaling solution depends on $\sigma$, and under which conditions the above bifurcation occurs. This is shown in Fig. \ref{fig7}. In order to understand this behaviour one should note that when $\sigma\rightarrow0$ (while still fulfilling the condition of Eq.~(\ref{cond1})) we have
\be
v_{s}=1\,,\quad \alpha=\frac{\sigma}{1+\lambda}\,,
\ee
that is, $\alpha\rightarrow0$. In the opposite limit, when $\sigma\rightarrow\infty$
\be
\alpha=\frac{\sigma}{1-\lambda}\,,\quad v_{s}=k\frac{1-\lambda}{2\lambda\sigma}\,,
\ee
so $\alpha\rightarrow\infty$ and $v_s\rightarrow0$. For the parameters we have been using ($k=1$ and $\lambda=2/3$) the latter equation yields 
\be
v_{s}=\frac{1}{4\sigma}\,.
\ee
It is noteworthy that the bifurcation into the two different branches (according to the initial conditions) only happens for an intermediate range of values of the parameter $\sigma$. Below this range the attractor solution is always the ultrarelativistic branch, while above it the attractor is the normal branch. 

\begin{figure}
\includegraphics[width=3.3in,keepaspectratio]{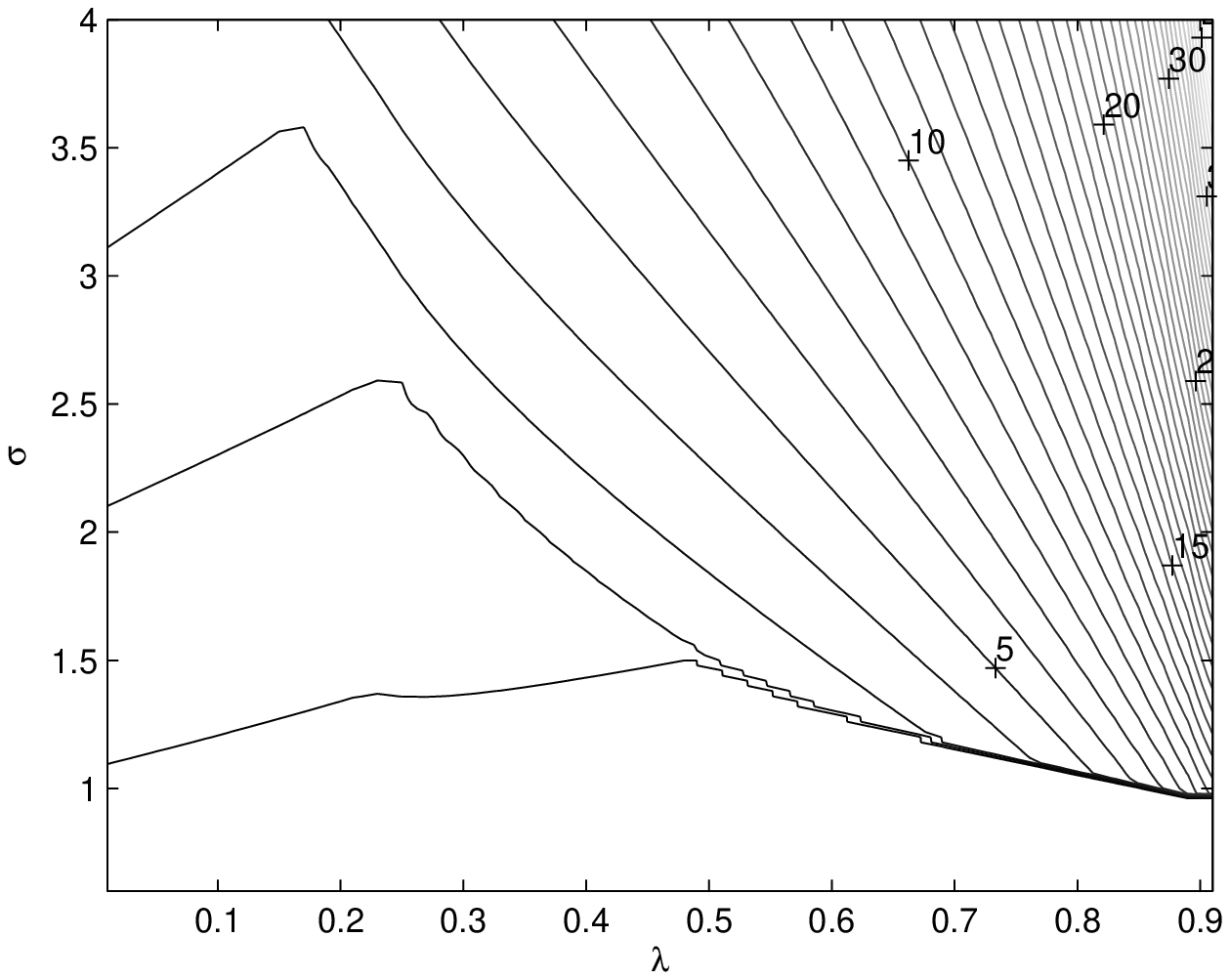}
\includegraphics[width=3.3in,keepaspectratio]{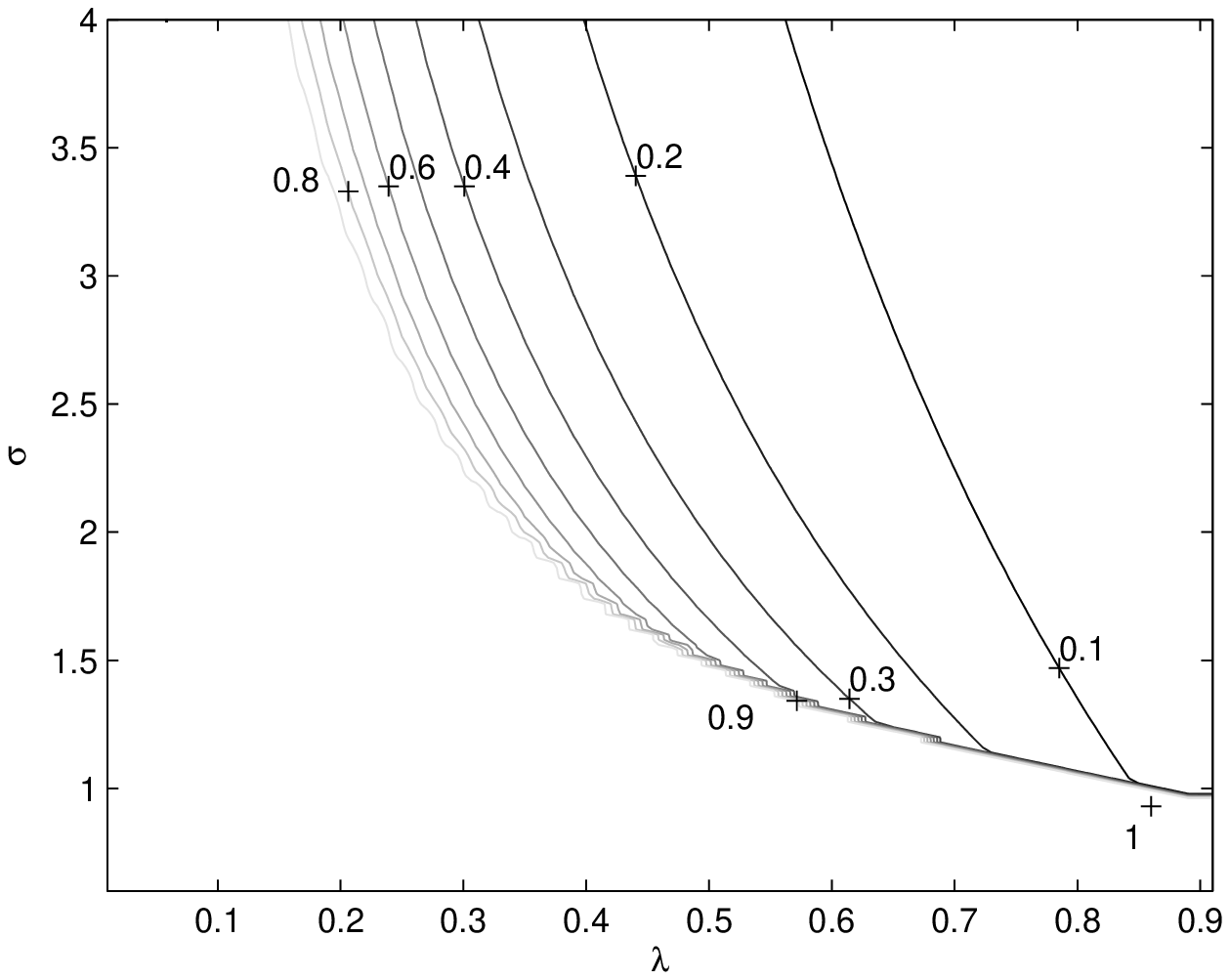}
\caption{\label{fig8}The scaling behaviour of $\alpha$ (top panel) and $v_s$ (bottom panel) as a function of $\sigma$ and the expansion rate $\lambda$, for a segment with initial parameters $\alpha_{i}=0.3$ and $v_{si}=0.7$; further parameter choices are $c_\star=0.23$ and $k=1$.}
\end{figure}

Finally, Fig. \ref{fig8} depicts the effect of different expansion rates on these attractor solutions. The asymptotic value of $\alpha$ increases significantly with $\lambda$, particularly for high values of $\sigma$. Recall that these solutions exist provided $\sigma>4\Gamma\gamma$ and that $\Gamma$ itself depends on $\lambda$. This threshold is more obvious for fast expansion rates, since this enhances the difference between the two solutions. Similarly for the velocities, the smaller the expansion rate the smaller will be the difference for values close to the threshold.

In conclusion, with this phenomenological term, initially small segments typically reach a scaling solution with smaller sizes and ultrarelativistic velocities, while large segments typically grow and reach nonrelativistic velocites. As a simple way to verify how robust (and model-independent) these results are, we will now consider an alternative form for this term.

\subsection{Balance Equation}

Here we will consider the following alternative form for the evolution equations
\be\label{ls_eqn_brown}
\frac{dl_{s}}{dt}=Hl_{s}-v_{s}^{2}\frac{l_{s}}{l_{d}}+\left(d\frac{v_{s}l_{s}}{L}-k_{1}\right)
\ee
\be
\frac{dv_{s}}{dt}=\left(1-v_{s}^{2}\right)\left[\frac{k}{l_{s}}-\frac{v_{s}}{l_{d}}\right]
\ee
In this we are assuming that the network of string segments has a Brownian distribution, something that can be tested in numerical simulations. The new term accounts for the probability that different segments intersect, which depends both on the length/number density and velocity of the segments. In this case there are two new free parameters, $d$ and $k_1$.

Before studying the scaling behaviour of this system, let us briefly describe how equation (\ref{ls_eqn_brown}) is derived.  Consider a monopole network with characteristic length $L$ and think of the monopoles as being connected by string segments of average length $L$. Now consider a string segment of length $l_s<L$ moving with velocity $v_s$ in this network.  Assuming $L$ has a Brownian structure, the probability that the segment $l_s$ will meet a segment of average length $L$ in time $\delta t$ is:
\be
\delta P = \frac{v_s \delta t}{L} \frac{l_s}{L}=\frac{v_s l_s \delta t}{L^2} \,.        
\ee
Soon after the segments meet, a monopole from one segment interacts with an anti-monopole from the other segment and annihilates, so the segment $l_s$ joins with $L$ increasing its length to $l_s+L\simeq L$. Considering now a collection of segments of average length $l_s$, so there are $L^3/l_s^3$ segments in a shell of volume $L^3$, the rate of change of network energy density $\rho_s$ due to such joining processes is: 
\be
\dot \rho_{s,{\rm join}} = - d \frac{v_s l_s}{L^2} \frac{\mu (l_s+L)}{L^3} 
\frac{L^3}{l_s^3} \simeq d \frac{v_s \mu}{l_s^2 L} \, ,
\ee     
where the coefficient $d$ quantifies the efficiency of this joining process. This term, when translated to a rate of change of length $l_s$ gives rise precisely to the term with coefficient $d$ in equation (\ref{ls_eqn_brown}).  The constant term $k_1$ describes the shrinking of the segment $l_s$ due to a monopole acceleration $f_m=k_1/l_s$ giving rise to 
\be
\dot \rho_{s,{\rm shrink}}=\frac{\mu k_1}{l_s^3} \,.
\ee

We can now search for scaling solutions, as in the previous case. Again we find two branches, an ultrarelativistic one
\be
v_{s}=1\,,\quad \alpha=\frac{k_{1}}{\frac{d}{\gamma}-(1+\lambda)},
\ee
subject to the condition that
\be
d>\gamma(1+\lambda)\,,
\ee
and a normal one
\be
\alpha=\frac{k_{1}}{\lambda(1-2v_{s}^{2})+d\frac{v_{s}}{\gamma}-1}
\ee
\be
v_{s}=\frac{k}{2\lambda\alpha}
\ee
which requires
\be
d>\frac{\gamma}{v_{s}}\left[1+\lambda\left(2v_{s}^{2}-1\right)\right]\,.
\ee
Obviously scaling disappears if
\be
k_{1}>d\frac{v_{s}l_{s}}{L}\,.
\ee

\begin{figure}
\includegraphics[width=3.3in,keepaspectratio]{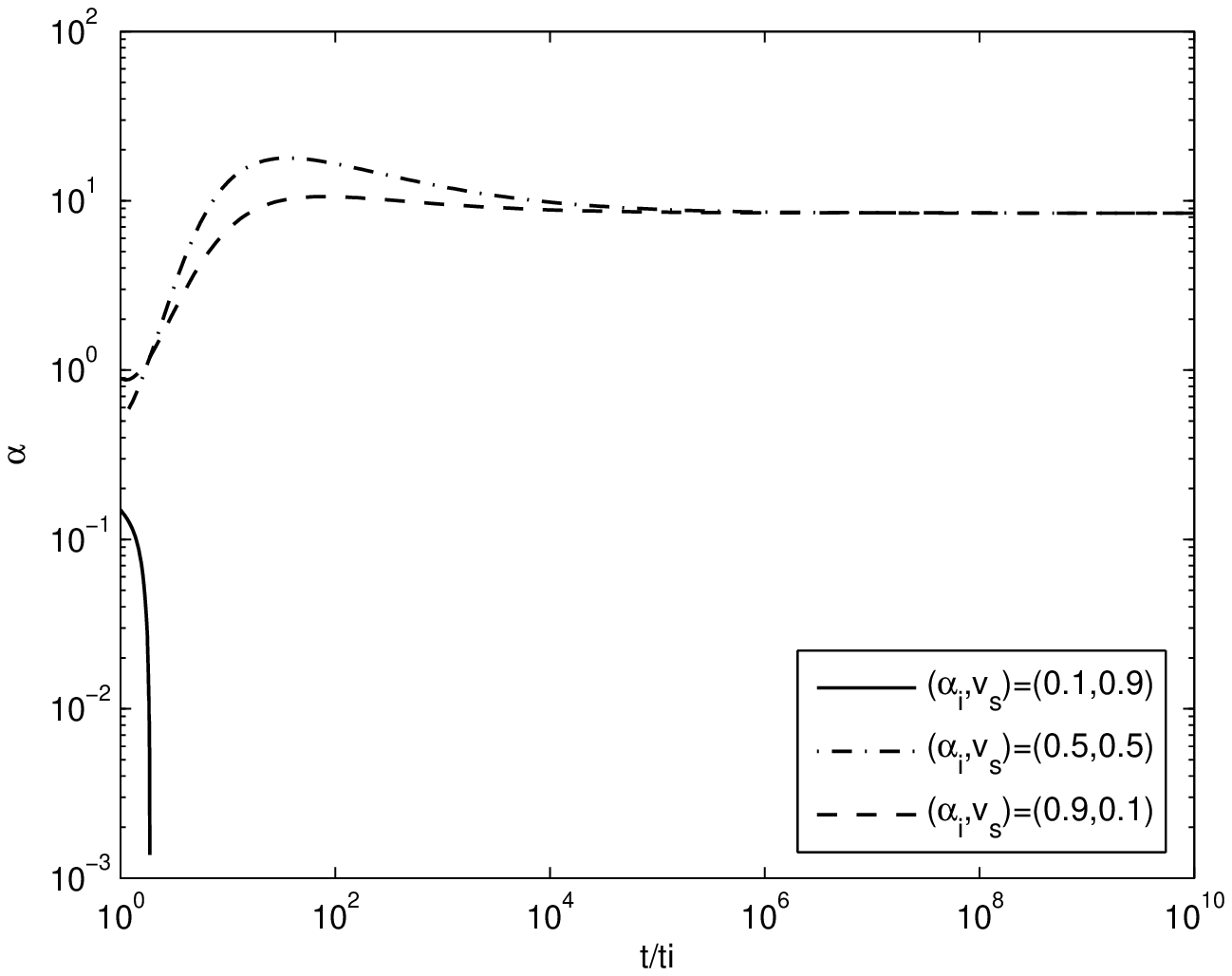}
\includegraphics[width=3.3in,keepaspectratio]{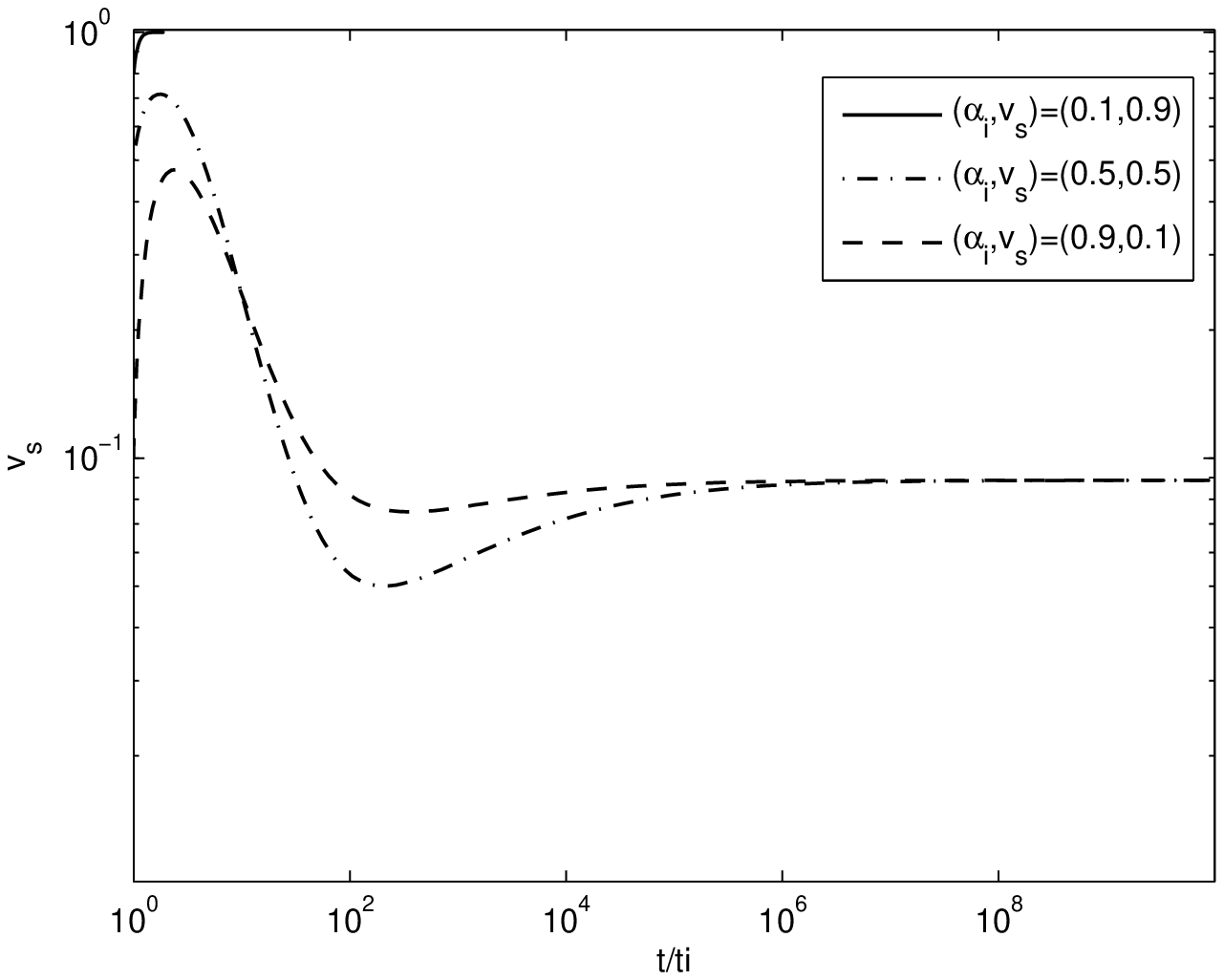}
\caption{\label{fig9}The time evolution $\alpha$ (top panel) and $v_s$ (bottom panel) for three different segments in the matter era; further parameter choices are $d=0.02$, $k_1=0.5$, $c_\star=0.23$ and $k=1$.}
\end{figure}

\begin{figure}
\includegraphics[width=3.3in,keepaspectratio]{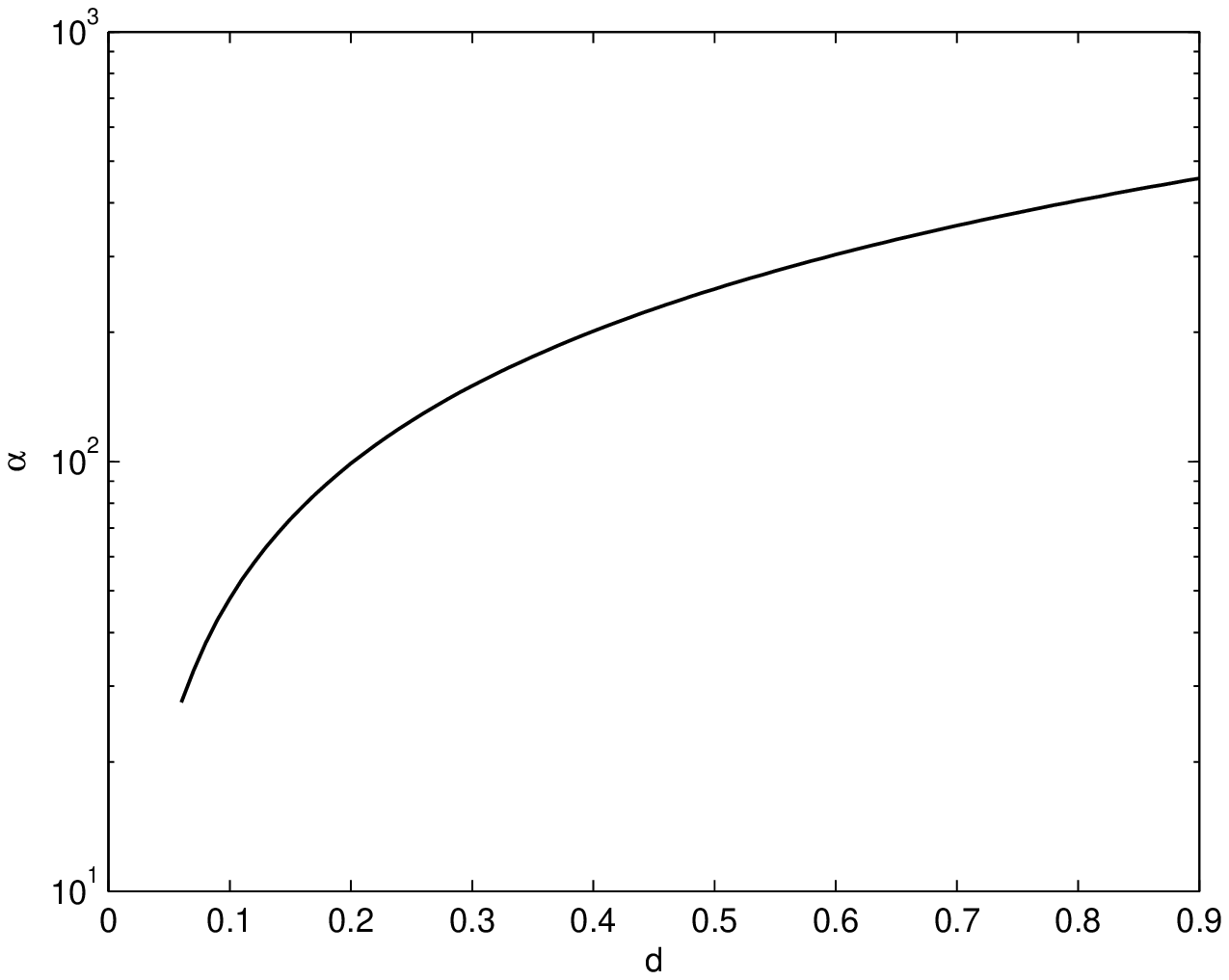}
\includegraphics[width=3.3in,keepaspectratio]{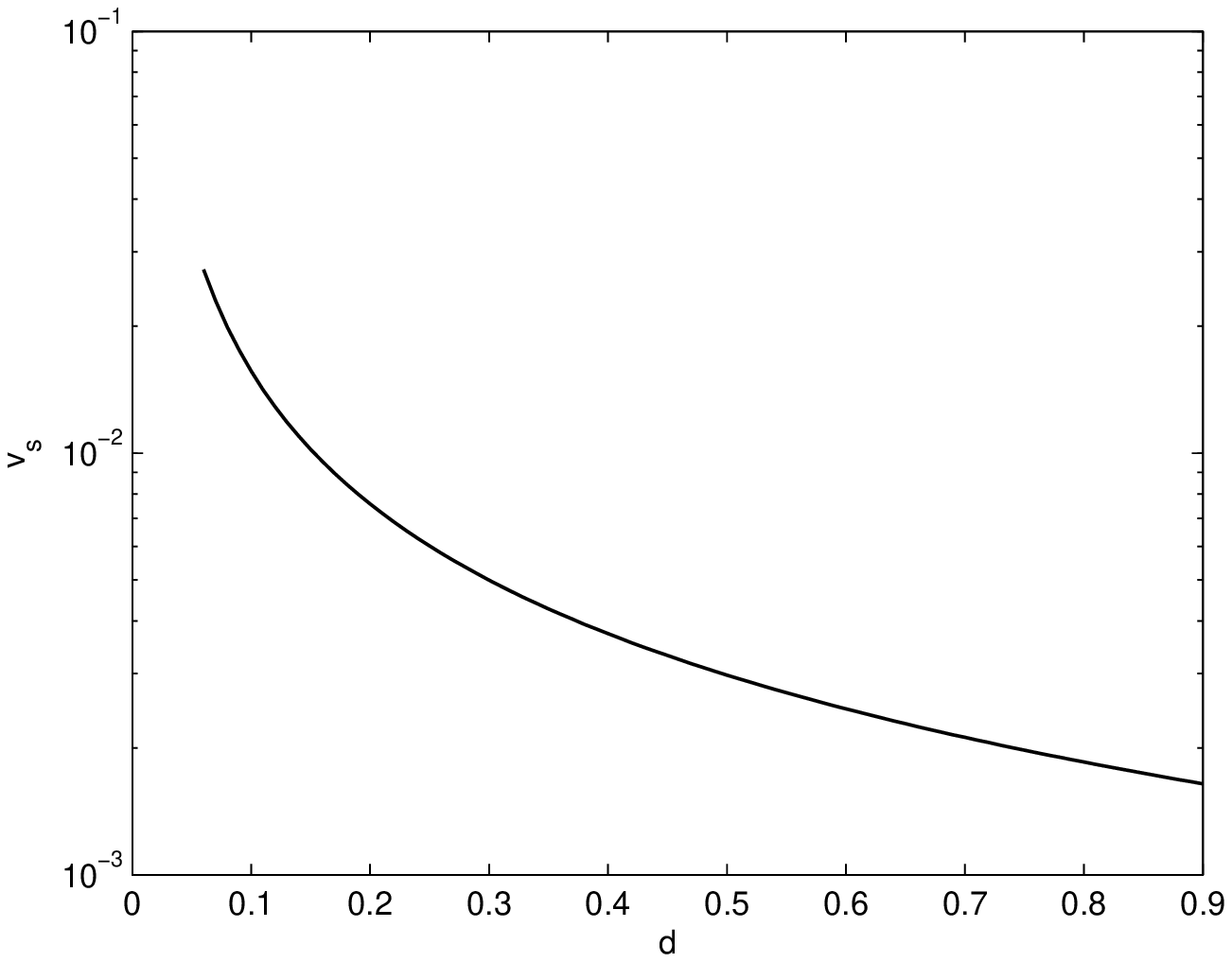}
\caption{\label{fig10}The asymptotic values of $\alpha$ (top panel) and $v_s$ (bottom panel) in the matter-dominated era, for $k_1=1$, $c_\star=0.23$ and $k=1$.}
\end{figure}

A matter era example of the evolution of the segments in this model is shown in Fig. \ref{fig9}. Here large enough segments reach scaling (for this particular choice of parameters this is $\alpha=8.461$, $v_{s}=0.08864$), while for small segments (for which the sum of the $d$ and $k_1$ terms is negative) there is no scaling at all, and the segments will shrink and disappear. Note that one can easily choose parameters such that $\alpha>>1$, in other words the characteristic size of the segments becomes much larger than the horizon and the segments become rare; this can be seen in Fig. \ref{fig10}.

\begin{figure}
\includegraphics[width=3.3in,keepaspectratio]{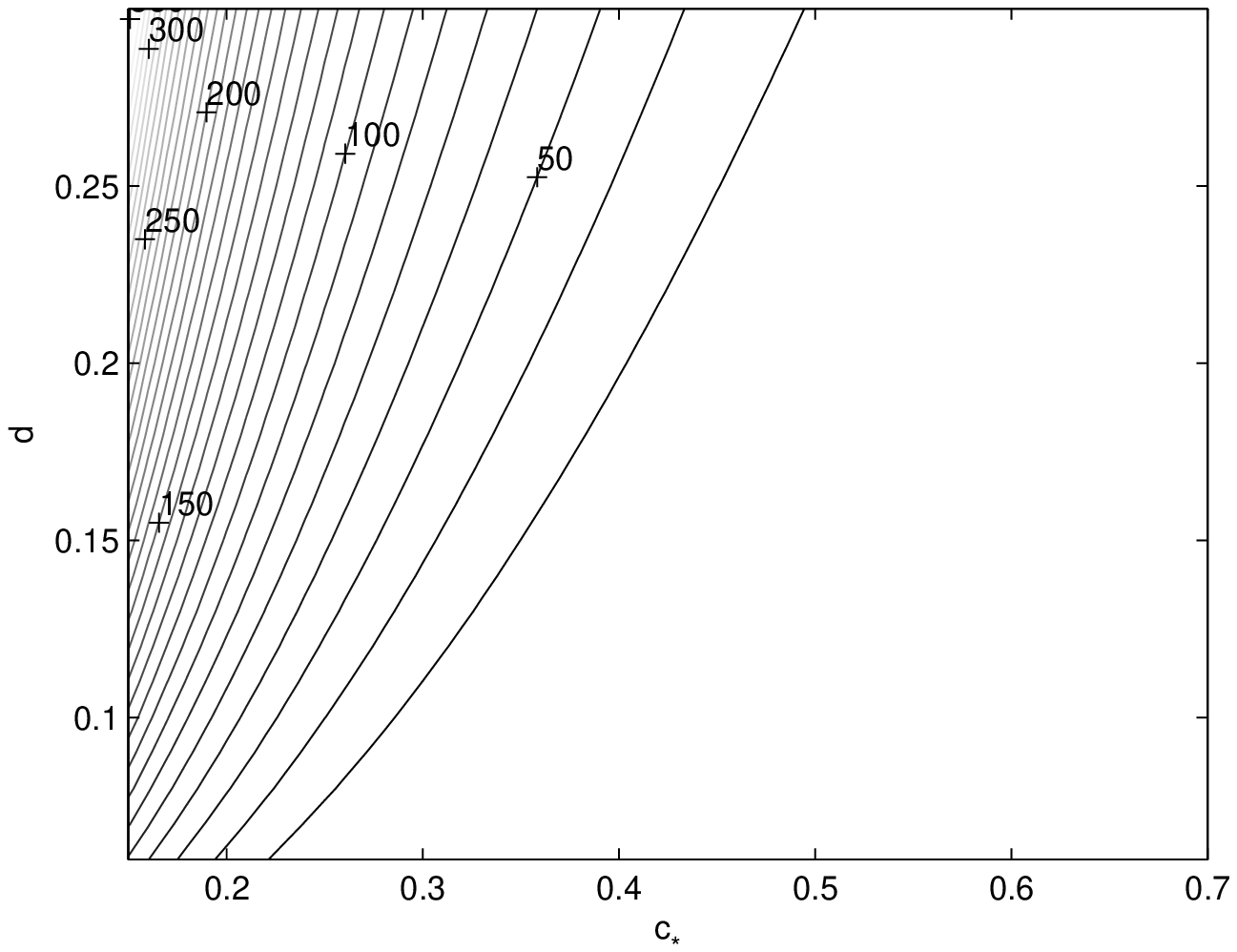}
\includegraphics[width=3.3in,keepaspectratio]{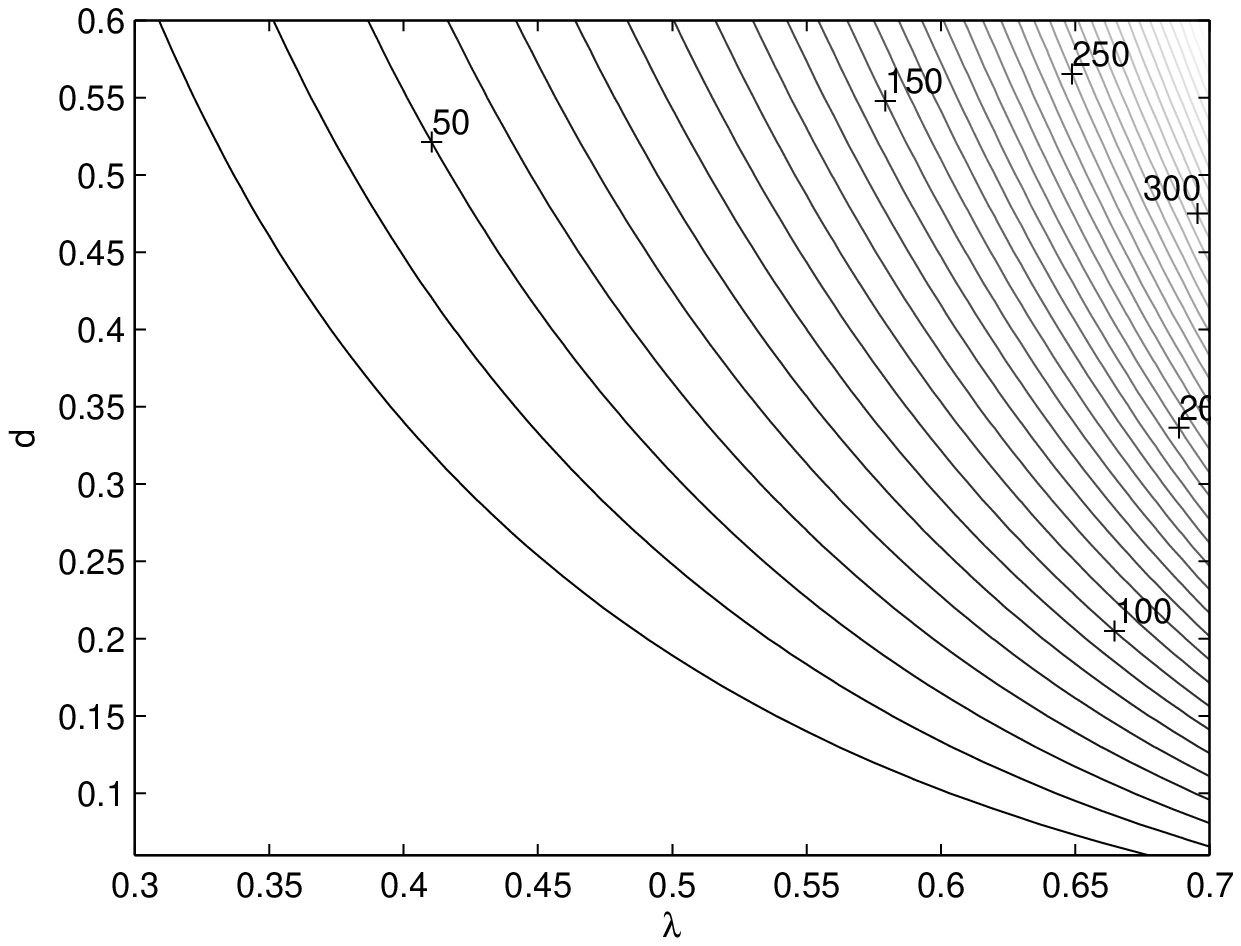}
\caption{\label{fig11}The asymptotic value of $\alpha$ as a function of $d$ and $c_\star$ (top panel) and of $d$ and $\lambda$ (bottom panel). In both cases the plots assumed $k_1=1$ and $k=1$; the top panel is for the matter era, while the bottom one has $c_\star=0.23$.}
\end{figure}

Finally, Fig. \ref{fig11} depicts the behaviour of the scaling solution (when it exists) as a function of several model parameters. As expected, a large $c_\star$ (implying a large monopole separation) or a slow expansion rate make the existence of a scaling attractor more difficult, and can eliminate it altogether. In those circumstances all segments would eventually disappear. This would then lead us to expect more and/or longer segments in the matter era than in the radiation one.

\section{Comparing with simulations}
\label{comp}

It is vital to compare the results obtained from the analytical approach with some numerical simulations. The (simplest) model for semilocal strings 
that can provide us with some numerical simulations for comparison is the one introduced in section~\ref{semi}. As explained there, only low values of $\beta$ are considered, because higher values render the strings unstable or make the network disappear very fast. The actual simulations used in this work for a preliminary comparison to our analytic models are the (existing) ones described in \cite{Achucarro:2007sp}. We refer the interested reader to that reference, and only proceed to sketch that approach in this section (focusing on the aspects that are directly relevant for us). 

In a $512^3$ box with periodic boundary conditions we discretize the equations of motion coming from Eq~(\ref{SLaction}).  As initial condition, we set up random velocities for the scalar fields in the semilocal model. After a transient time, the system forms a network that approaches a scaling regime. We are not interested in capturing physical details during the phase transition; instead, we are interested in the long term properties of the network. This is why we only monitor the simulation to detect semilocal string segments once the system has relaxed to the scaling regime.

\begin{table}
\begin{center}
\begin{tabular}{l|c|c|c|c|c}
\hline
\hline
Simulations & Time & ${\cal L}$ & ${\cal N}$ & $\gamma_s$ & $\gamma_m$ \\
\hline
box=$512^3$ & $150$ & $18821.8 \pm 378.8$ & $462.4 \pm 15.2$ & 0.56 & 0.44 \\
$\delta=0.02$ & $200$ & $15462.9 \pm 314.9$ & $307.3 \pm 14.0$ & 0.47 & 0.40 \\
$\beta=0.04$ & $250$ & $13426.6 \pm 357.6$ & $222.4 \pm 9.3$ & 0.40 & 0.34 \\
$N=9$ & $300$ & $11880.0 \pm 423.2$ & $167.8 \pm 9.4$ & 0.35 & 0.31 \\
\hline
box=$512^3$ & $150$ & $19811.3 \pm 529.2$ & $530.5 \pm 12.8$ & 0.55 & 0.42 \\
$\delta=0.02$ & $200$ & $16214.8 \pm 293.6$ & $346.5 \pm 14.4$ & 0.45 & 0.36 \\
$\beta=0.06$ & $250$ & $14078.5 \pm 110.1$ & $241.5 \pm 7.4$ & 0.39 & 0.33 \\
$N=6$ & $300$ & $12521.4 \pm 142.6$ & $183.5 \pm 10.6$ & 0.35 & 0.30 \\
\hline
box=$512^3$ & $150$ & $13535.3 \pm 362.4$ & $192.4 \pm 9.6$ & 0.66 & 0.60 \\
$\delta=0.05$ & $200$ & $9888.8 \pm 360.6$ & $103.1 \pm 10.2$ & 0.58 & 0.55 \\
$\beta=0.04$ & $250$ & $7624.3 \pm 348.5$ & $56.4 \pm 6.7$ & 0.53 & 0.53 \\
$N=16$ & $300$ & $6254.0 \pm 346.2$ & $36.9 \pm 5.9$ & 0.48 & 0.51 \\
\hline
box=$512^3$ & $150$ & $12224.5 \pm 345.4$ & $218.8 \pm 13.1$ & 0.70 & 0.57 \\
$\delta=0.05$ & $200$ & $9000.5 \pm 482.5$ & $111.7 \pm 9.3$ & 0.61 & 0.53 \\
$\beta=0.06$ & $250$ & $7111.1 \pm 558.4$ & $62.4 \pm 8.5$ & 0.55 & 0.52 \\
$N=23$ & $300$ & $5953.2 \pm 572.4$ & $43.5 \pm 7.4$ & 0.50 & 0.49 \\
\hline
box=$512^3$ & $150$ & $14421.9 \pm 382.7$ & $244.5 \pm 19.3$ & 0.64 & 0.55 \\
$\delta=6/t$ & $200$ & $10166.8 \pm 398.6$ & $137.7 \pm 27.1$ & 0.57 & 0.50 \\
$\beta=0.04$ & $250$ & $7499.6 \pm 425.1$ & $93.1 \pm 27.3$ & 0.54 & 0.45 \\
$N=16$ & $300$ & $5825.6 \pm 455.1$ & $74.7 \pm 24.8$ & 0.51 & 0.41 \\
\hline
box=$512^3$ & $150$ & $12980.1 \pm 196.2$ & $263.2 \pm 14.1$ & 0.68 & 0.53 \\
$\delta=6/t$ & $200$ & $9155.0 \pm 224.0$ & $124.7 \pm 11.3$ & 0.61 & 0.51 \\
$\beta=0.06$ & $250$ & $6813.2 \pm 336.9$ & $65.8 \pm 7.9$ & 0.56 & 0.51 \\
$N=12$ & $300$ & $5413.9 \pm 327.6$ & $44.0 \pm 8.2$ & 0.52 & 0.48 \\
\hline
\hline
\end{tabular}
\caption{The measured values (with statistical errors only) of the total string length ${\cal L}$ and monopole number ${\cal N}$ for the various series of simulations further described in the text, and the inferred (approximate values of the string and monopole scaling parameters, $\gamma_s$ and $\gamma_m$.}
\label{tab1}
\end{center}
\end{table}

Unlike ordinary cosmic strings, semilocal strings are not topological and that makes them more complicated to track on a lattice simulation. It can be the case that even though semilocal strings carry magnetic field strength, they have no zeroes of the scalar field in their core, so tracking zeros of scalar fields is not the best approach to  detect the strings. The strategy used here is the one following \cite{Urrestilla:2001dd,Pickles:2002ym,Achucarro:2007sp}: if the magnetic field strength at a given point exceeds a certain fraction of the maximum magnetic field of the corresponding Abelian Higgs cosmic string core, the point is considered to be part of a semilocal string. This would give us the `volume' of the string, which we turn into a length dividing by the width of the corresponding Abelian Higgs string. Segments that are too short (typically shorter than a couple of times the width of the strings) are discarded.  

Therefore, we take snapshots of the simulated volume at several time steps well within the scaling regime, search for lattice points which have significant magnetic field strength, compute how those points are collected into string segments, and calculate their length. It is the number density and length density of those strings that we use to compare to the predictions of the analytic models put forward in this paper. In what follows we do not directly compare the velocities in the model and simulations, since reliable numerical measurements of these velocities are highly non-trivial and require the development of additional numerical algorithms, which must be left for future work. For analogous issues in the more standard case of Abelian-Higgs string networks, see \cite{Moore}; for the case of domain wall simulations with the Press-Ryden-Spergel algorithm see \cite{WallSim}.

\begin{figure}
\includegraphics[width=3.3in,keepaspectratio]{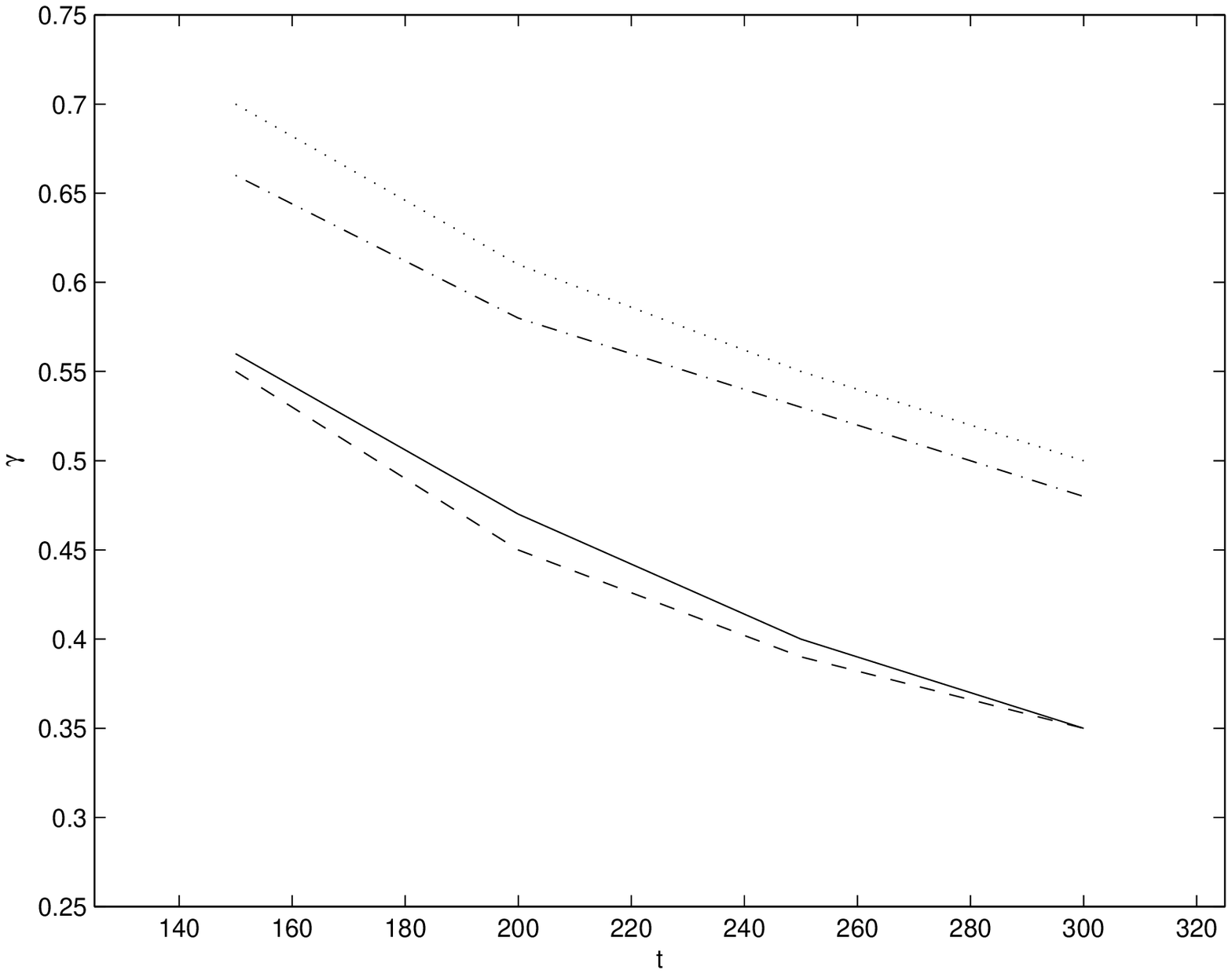}
\includegraphics[width=3.3in,keepaspectratio]{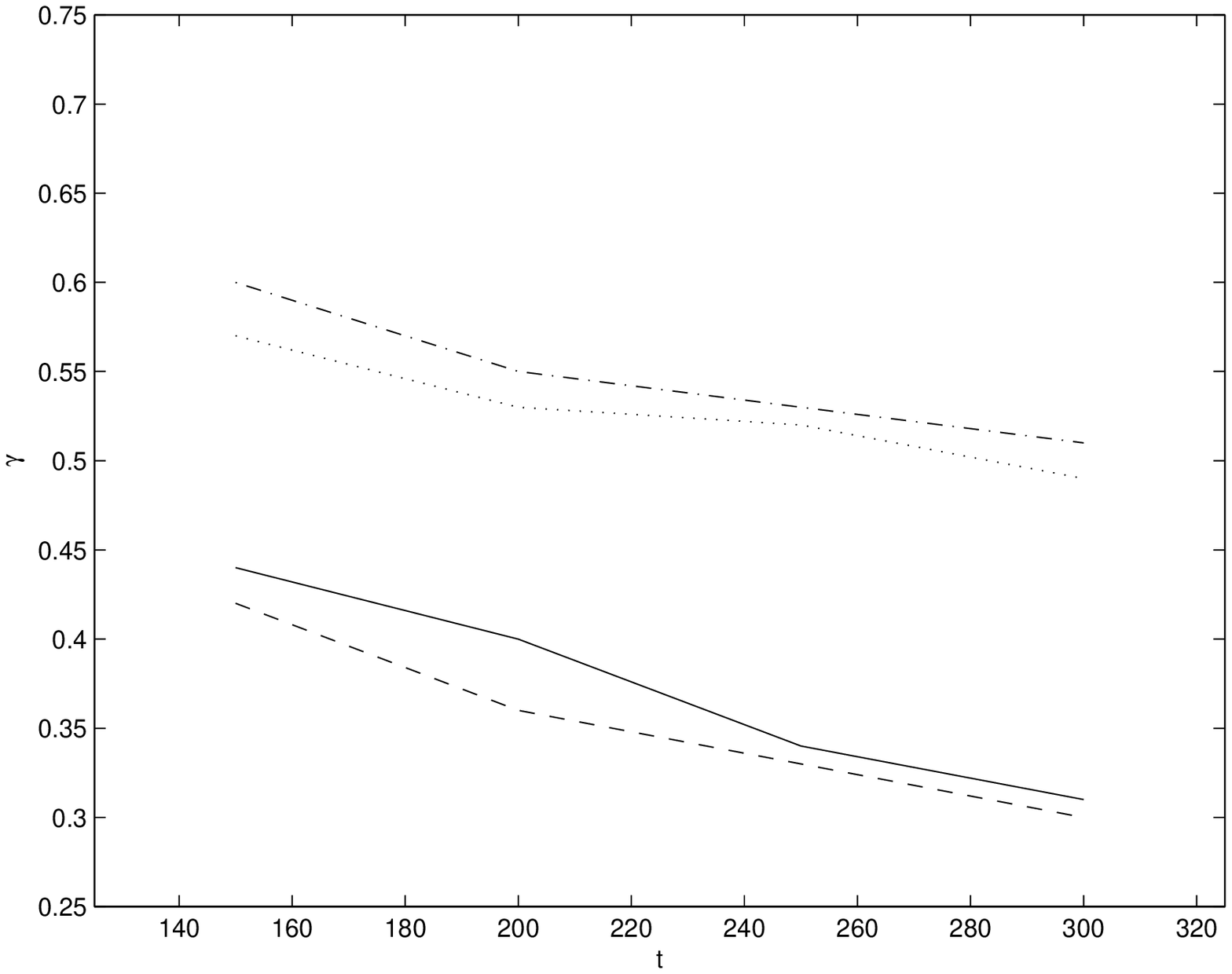}
\caption{\label{fig12}The evolution of the string scaling parameter $\gamma_s=L_s/t$ (top panel) and the monopole scaling parameter $\gamma_m=L_m/t$ (bottom panel), for four different series of constant-damping simulations. In each panel the four lines correspond to the following cases: $\delta=0.02$, $\beta=0.04$ (solid), $\delta=0.02$, $\beta=0.06$ (dashed), $\delta=0.05$, $\beta=0.04$ (dash-dotted), $\delta=0.05$, $\beta=0.06$ (dotted). Further details are provided in the main text.}
\end{figure}

\begin{figure}
\includegraphics[width=3.3in,keepaspectratio]{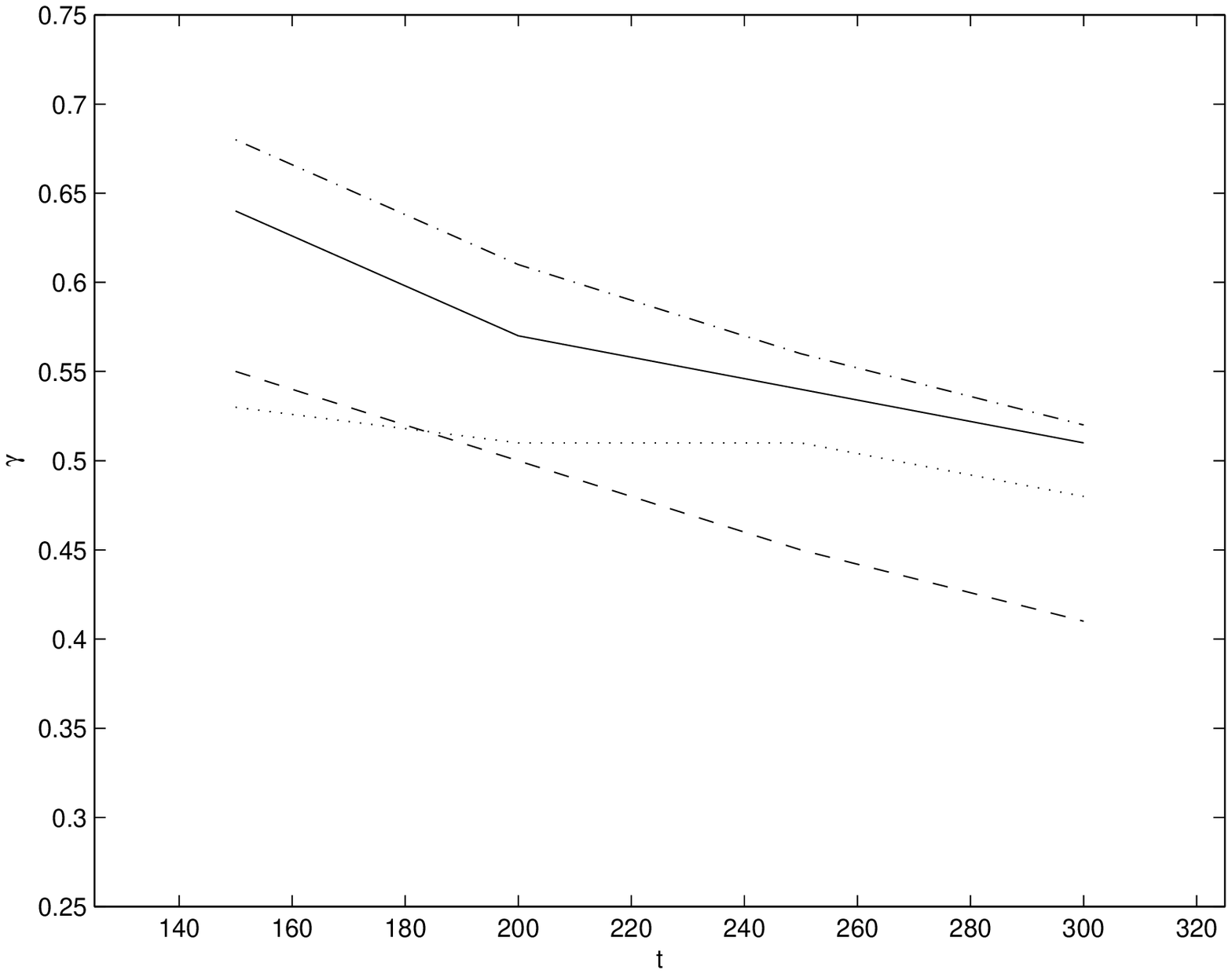}
\caption{\label{fig13}The evolution of the string and monopole scaling parameters for two different series of simulations with a damping coefficient $\delta=6/t$. The solid and dashed lines respectively depict $\gamma_s$ and $\gamma_m$ for simulations with $\beta=0.04$; the dash-dotted and dotted ones show the same quantities for $\beta=0.06$ simulations. Further details are provided in the main text.}
\end{figure}

We used results from several existing sets of simulations (of the order of 10 simulations for each value of the coupling $\beta$ and the damping term $\delta$) in order to obtain basic statistics about the properties of the networks (details of the simulations can be found in Table~\ref{tab1}). The fact that these are `legacy' simulations (not done with the present purpose in mind) explains the difference in the number of simulations in each set, and also the fact that most of them have a constant damping term rather than a decaying one (which would more accurately describe an expanding universe). Nevertheless, these are sufficient for our present purposes.

As mentioned before, we are not focusing on the first time steps of the simulations, and the code was tuned to get into the scaling regime as fast as possible. Once in the scaling regime, the network was evolved using different (constant or time-decaying) damping terms. For each series we follow the procedure described above, which gives us, for every simulation, a good approximation to the total string length ${\cal L}$ and monopole number ${\cal N}$ in the box. We proceed to average those two numbers over the various realisations, for several late-time boxes. The number of monopoles is obtained by counting the number of segments and multiplying by two, since each segment has one monopole at each end. These will slightly overcount monopoles, since it does not take into account segments that have closed to form a loop. There are other systematic errors as well in this `number counting' approach. Therefore one should realise that the errors quoted are only statistical errors, and provide only a lower bound of the uncertainties in these simulations. 

We then translate the obtained string lengths and number of monopoles into a VOS-type lengthscale, which we calculate in two independent ways, either using the strings or the monopoles:
\be
\gamma_s\equiv\frac{L_s}{t}=\frac{1}{t}\sqrt{\frac{V}{\cal L}}\,,\quad \gamma_m\equiv\frac{L_m}{t}=\frac{1}{t}\left(\frac{V}{\cal N}\right)^{1/3}\,.
\ee
Given the way they were calculated, $L_s$ should be thought of as the typical inter-string distance (or perhaps the typical segment size), while $L_m$ is a characteristic inter-monopole distance. These are therefore not correlation lengths in the same strict sense as the term is used, for example, in Goto-Nambu string simulations. In particular, the fractal distribution of the semilocal networks (and more specifically the assumption of a Brownian network) is an issue that warrants further study. Given the possible systematic uncertainties associated with the calculations of ${\cal L}$ and ${\cal N}$, we decided to present $\gamma_s$ and $\gamma_m$ without error bars. Figures \ref{fig12} and \ref{fig13} summarise our results.

We start by noting the remarkable similarity between $\gamma_s$ and $\gamma_m$. The agreement is even more striking if one observes that, due to the previously mentioned possibility of an overcounting bias, the $\gamma_m$ we obtain this way should be somewhat smaller than the correct one. Although we can't quantify the magnitude of this effect, inspection of animations of the simulations indicates that the number of loops is relatively small, and thus so should be the overcounting. Given the underlying uncertainties the two results can therefore be said to be in full agreement.

We also find excellent agreement in the results of the simulations which only differ in the value of $\beta$; therefore the two available values of $\beta$ are too close to each other to be distinguishable by this method. On the other hand, the effect of the different damping terms is quite obvious: a larger damping reduces the string length and the number of monopoles, thus leading to a larger $\gamma$. While keeping in mind the aforementioned uncertainties, we note that our results for the constant damping cases are consistent with the asymptotic value of the energy density in the network being inversely proportional to the damping, in other words $\rho\propto\delta^{-1}$. This is to be expected, for example if one considers the energy losses of a given length of string moving subject to constant damping. One may also expect the string and monopole velocities to decrease as the damping increases, but in this case the exact behaviour is not easy to determine, and one may also expect a noticeable effect of the parameter $\beta$.

Finally, the absence of information on the velocities prevents us from providing a direct calibration of the parameters of the analytic model. Nevertheless, we can use the results of the time-decaying damping case to attempt a naive back-of-the-envelope exercise. Referring to Eqs. (\ref{nscaling1}-\ref{nscaling2}) we note that we have in principle three unknowns (the model parameters $c_\star$ and $k_m$, plus the velocity $v_0$). We then note that the damping term $\delta=6/t$ (where in an expanding universe context this $t$ of the simulations is effectively the conformal time) corresponds to a pysical-time expansion rate $a\propto t^{\lambda}$ with $\lambda=3/4$. If we now use $\gamma\sim0.5$ in Eqs. (\ref{nscaling1}-\ref{nscaling2}), we find the approximate relations
\be
c_\star k_m\sim 3.2\,
\ee
and
\be
v_0\sim0.12 k_m\,.
\ee
These are plausible values. In the case of Goto-Nambu strings both the energy loss parameter $c$ and the curvature paramater $k$ are of order (but slightly smaller than) unity \cite{VOS02}. Here one may expect a slightly larger $c_\star$ since the energy density of semilocal strings is somewhat smaller than that of Goto-Nambu ones: in the context of the analytic model, this would primarily be controlled by the parameter $c_\star$. As for our estimated velocity, this is somewhat lower than the ones typically encountered in other field theory defect simulations \cite{Moore,WallSim}, but one should also keep in mind that these previous measurements have almost always been performed in the matter or radiation eras---in other words, they have slower expansion rates, and therefore an effectively smaller damping term.

In summary, although the above comparison is quite simplistic, the results are at least encouraging. As has already been pointed out, a more detailed comparison and a proper calibration of the analytic models must wait for the numerical implementation of a reliable method to measure defect velocities in semilocal string simulations.

\section{Conclusions}

We have extended previously developed analytic models for defect evolution, applying them to the study of semilocal string networks. We have shown that a linear scaling evolution (analogous to the well-known one for cosmic strings) is the attractor solution for a broad range of model parameters. Our results therefore provide supporting evidence for those of existing numerical simulations, where confirmation of the presence of this solution is restricted to the limited dynamical range available.

We have discussed in some detail the evolution of individual semilocal string segments, focusing on the interesting phenomenology of segment growth and considering two different possible scenarios thereof. As expected (and as seen in simulations) we find that small segments tend to shrink and decay while large ones can grow and merge with others. Which (if any) of the two scenarios we have considered is the valid one is an issue that can only be resolved with future and more detailed simulations.

Finally we have used the results of existing simulations in order to obtain a very preliminary comparison with our analytic results. This endeavour is quite limited by the fact that the available simulations have not measured the defect velocities, which play a key role in VOS-type models. We found some encouraging results, but it is quite clear that a more detailed comparison, using expanding-universe simulations that explore a wider range of the space of model parameters (notably the coupling parameter $\beta$), will be needed in order to provide a meaningful calibration of the model. We shall address these issues in a subsequent publication.

\begin{acknowledgments}
This work was done in the context of the cooperation grant `Evolution and Astrophysical Consequences of Cosmic Strings and Superstrings' (ref. B-13/10), funded by CRUP and The British Council. We acknowledge stimulating discussions on the subject of this work with Ana Ach\'ucarro and Mafalda Leite.

A.A. is supported by the Centre for Theoretical Cosmology (CTC) in DAMTP, Cambridge, through a CTC Fellowship.
The work of C.M. is funded by a Ci\^encia2007 Research Contract, funded by FCT/MCTES (Portugal) and POPH/FSE (EC). The work of A.N. was partially funded by grant CAUP-10/2009-BII. A.N., A.A. and C.M. also acknowledge additional support from project PTDC/FIS/111725/2009 from FCT, Portugal. J.U. acknowledges financial support from the Basque Government (IT-559-10), the Spanish Ministry (FPA2009-10612), and the Spanish Consolider-Ingenio 2010 Programme CPAN (CSD2007-00042).

We acknowledge the allocation of computing time on the UK National Cosmology Supercomputer funded by PPARC, HEFCE and Silicon Graphics.
\end{acknowledgments}

\bibliography{semilocal}
\end{document}